%% file: main.tex
\documentclass[sigconf, nonacm]{acmart}

\newcommand\vldbdoi{10.14778/3489496.3489509}
\newcommand\vldbpages{285 - 298}
\newcommand\vldbvolume{15}
\newcommand\vldbissue{2}
\newcommand\vldbyear{2022}
\newcommand\vldbauthors{\authors}
\newcommand\vldbtitle{\shorttitle} 
\newcommand\vldbavailabilityurl{https://github.com/yushangdi/parChain}
\newcommand\vldbpagestyle{empty}

\usepackage{style}
\input{macro}

\begin{document}
\title{\framework: A Framework for Parallel Hierarchical Agglomerative Clustering using Nearest-Neighbor Chain}


\author{Shangdi Yu}
\affiliation{\institution{MIT CSAIL}}
\email{shangdiy@mit.edu}
\author{Yiqiu Wang}
\affiliation{\institution{MIT CSAIL}}
\email{yiqiuw@mit.edu}
\author{Yan Gu}
\affiliation{\institution{UC Riverside}}
\email{ygu@cs.ucr.edu}
\author{Laxman Dhulipala}
\affiliation{\institution{MIT CSAIL}}
\email{laxman@mit.edu}
\author{Julian Shun}
\affiliation{\institution{MIT CSAIL}}
\email{jshun@mit.edu}

\input{abstract}

\maketitle

\ifarxiv
\else
\pagestyle{\vldbpagestyle}
\begingroup\small\noindent\raggedright\textbf{PVLDB Reference Format:}\\
\vldbauthors. \vldbtitle. PVLDB, \vldbvolume(\vldbissue): \vldbpages, \vldbyear.\\
\href{https://doi.org/\vldbdoi}{doi:\vldbdoi}
\endgroup
\begingroup
\renewcommand\thefootnote{}\footnote{\noindent
This work is licensed under the Creative Commons BY-NC-ND 4.0 International License. Visit \url{https://creativecommons.org/licenses/by-nc-nd/4.0/} to view a copy of this license. For any use beyond those covered by this license, obtain permission by emailing \href{mailto:info@vldb.org}{info@vldb.org}. Copyright is held by the owner/author(s). Publication rights licensed to the VLDB Endowment. \\
\raggedright Proceedings of the VLDB Endowment, Vol. \vldbvolume, No. \vldbissue\ %
ISSN 2150-8097. \\
\href{https://doi.org/\vldbdoi}{doi:\vldbdoi} \\
}\addtocounter{footnote}{-1}\endgroup

\ifdefempty{\vldbavailabilityurl}{}{
\vspace{.3cm}
\begingroup\small\noindent\raggedright\textbf{PVLDB Artifact Availability:}\\
The source code, data, and/or other artifacts have been made available at \url{https://github.com/yushangdi/parChain}.
\endgroup
}
\fi

\input{fig_nnchain}
\input{intro}

\input{background}

\input{framework}
\input{nn_finding}

\input{cache}

\input{exp}
\input{related_work}

\input{conclusion}

\begin{acks}
This research was supported by DOE
Early Career Award\\
\#DESC0018947, NSF CAREER Award 
\#CCF-1845763, NSF Award \#CCF-2103483, Google Faculty Research Award, Google Research Scholar Award, DARPA SDH Award \#HR0011-18-3-0007, and Applications Driving Architectures (ADA) Research Center, a JUMP Center co-sponsored by SRC and DARPA.
\end{acks}


\bibliographystyle{ACM-Reference-Format}
\bibliography{ref}

\end{document}

%% file: macro.tex
\SetKwInput{KwRequire}{Prereq}
\SetKwInput{KwData}{Input}
\SetKwInput{KwResult}{Output}
\SetKwFor{PFor}{par\_for }{do}{end}
\newcommand{\set}[1]{\{#1\}}

\newcommand{\distcomp}{\mathfrak{D}\xspace}

\newcommand{\chain}{\mathcal{L}\xspace}
\newcommand{\terminal}{\mathcal{Z}\xspace}
\newcommand{\activeclusters}{\mathcal{A}\xspace}
\newcommand{\clusters}{\mathcal{A}\xspace}
\newcommand{\cache}{\mathcal{H}\xspace}
\newcommand{\edges}{\mathcal{E}\xspace}

\newcommand{\hierarchy}{\mathcal{T}_H\xspace}
\newcommand{\cl}{C}
\newcommand{\cld}{\Delta}

\newcommand{\Var}{\text{Var}\xspace}

\newcommand{\tree}{\mathcal{T}}
\newcommand{\kdt}{$k$d-tree\xspace}

\newcommand{\nocid}{\textsc{NULL}\xspace}
\newcommand{\cid}{\textsc{cid}\xspace}

\newcommand{\writemin}{\textsc{WriteMin}\xspace}
\newcommand{\entry}[3]{\set{{#1}, {#2}}}

\newcommand{\termi}{terminal node\xspace}
\newcommand{\termis}{terminal nodes\xspace}
\newcommand{\rnn}{R-NN\xspace}
\newcommand{\rnns}{R-NNs\xspace}
\newcommand{\nnc}{NNC\xspace}

\newcommand{\framework}{ParChain\xspace}

\newcommand{\smb}[1]{{\scriptsize \mbox{\emph{#1}}}}

\newcommand{\myparagraph}[1]{\vspace{1pt} \noindent {\bf #1.}}
\newcommand{\defn}[1]{\emph{\textbf{#1}}}

\definecolor{myblue}{RGB}{0,128,255}

\newcommand{\custo}[1]{{\color{myblue} #1}}

\newif\ifarxiv
\arxivtrue

%% file: abstract.tex
\begin{abstract}
This paper studies the hierarchical clustering problem, where the goal
is to produce a dendrogram that represents clusters at varying scales
of a data set.  We propose the \framework framework for designing
parallel hierarchical agglomerative clustering (HAC) algorithms, and
using the framework we obtain novel parallel algorithms for the
complete linkage, average linkage, and Ward's linkage
criteria. Compared to most previous parallel HAC algorithms, which
require quadratic memory, our new algorithms require only linear
memory, and are scalable to large data sets. \framework is based on
our parallelization of the nearest-neighbor chain algorithm, and
enables multiple clusters to be merged on every round.  We introduce
two key optimizations that are critical for efficiency: a range query
optimization that reduces the number of distance computations required
when finding nearest neighbors of clusters, and a caching optimization
that stores a subset of previously computed distances, which are likely
to be reused.

Experimentally, we show that our highly-optimized implementations
using 48 cores with two-way hyper-threading achieve 5.8--110.1x speedup over
state-of-the-art parallel HAC algorithms and achieve 13.75--54.23x
self-relative speedup. Compared to state-of-the-art algorithms, our
algorithms require up to 237.3x less space. Our algorithms are able to scale
to data set sizes with tens of millions of points, which existing
algorithms are not able to handle.

\vspace{-7pt}
\end{abstract}

%% file: fig_nnchain.tex
\begin{figure*}[t!]
\begin{center}
\includegraphics[width=0.9\textwidth]{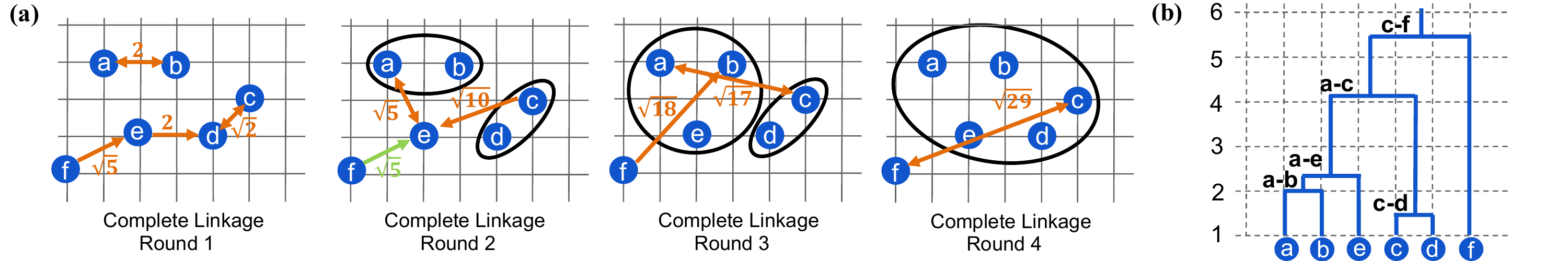}
\caption{\textbf{(a)} The four rounds of the nearest-neighbor chain algorithm on a six point data set using the complete linkage metric. The black circles are clusters containing more than one point. An arrow from point $x$ to point $y$ means that $x$'s cluster's nearest neighbor is $y$'s cluster.
  The orange arrows are new neighbors found on this round; the green arrow means the nearest neighbor is not updated on this round due to the reducibility property. 
  On Round 2, we avoided a nearest neighbor search for cluster $\set{f}$.
  The numbers on the arrows between pairs of points are the distances between the clusters that the points belong to according to the complete linkage metric (furthest distance between a pair of points, one from each cluster). Ties are broken lexicographically. 
  \textbf{(b)} The dendrogram for complete linkage clustering.
The label on each internal node corresponds to the furthest point pair in the two clusters that are merged in the algorithm, and its distance is equal to the node's height in the dendrogram.
}
\label{fig:nnchain}
\end{center}
\end{figure*}

%% file: intro.tex
\section{Introduction}\label{sec:intro}
Clustering is an unsupervised machine learning method that has been
widely used in many fields including computational biology, computer
vision, and finance to discover structures in a data
set~\cite{irbook,
  berkhin2006survey,eisen1998cluster,Aggarwal2013,Jain1999,Xu2015,leibe2006efficient,tumminello2010correlation}.
To group similar objects at all resolutions, a \emph{hierarchical
  clustering} can be used to produce a tree that represents
clustering results at different scales.
The resulting hierarchical cluster structure is called a
\textit{dendrogram}, which is a tree representing the agglomeration of
clusters, as shown in Figure~\ref{fig:nnchain}(b).

There is a rich literature on designing hierarchical agglomerative
clustering (HAC) algorithms~\cite{irbook}.  Unfortunately, exact HAC
algorithms usually require $\Omega(n^2)$ work,
since the distances between all
pairs of points have to be computed.
To accelerate exact HAC algorithms due to their significant
computational cost, there have been several parallel exact HAC
algorithms proposed in the
literature~\cite{olson1995parallel,jeon2014multi,li1990parallel,li1989parallel,du2005novel,zhang2019dhc,hung2014fast_protein_cluster,wang2021fast},
but most of them maintain a distance matrix, which requires quadratic
memory, making them unscalable to large data sets.
The only parallel exact algorithm that works for the metrics that we consider and uses
subquadratic space is by Zhang
et al.~\cite{zhang2019dhc}, but it has not been shown to scale to
large data sets.
In this paper, we propose the \framework framework for designing parallel exact HAC
algorithms that use \emph{linear memory}, based on the classic
nearest-neighbor chain algorithm.

The nearest-neighbor chain (\nnc) algorithm~\cite{mamano2019new} is a
popular algorithm that can be used for a wide range of HAC
metrics~\cite{benzecri1982construction,de1980classification,juan1982programme,
  murtagh1983survey, gronau2007optimal}. 
A \defn{nearest-neighbor chain  (\nnc)} is a linked list of nodes, where each node represents a cluster
and all except at most one node have a pointer to its nearest neighbor (its successor). The chain can start from an arbitrary cluster.
If a node does not have a pointer, its nearest neighbor is not yet computed, and this node is called a \defn{\termi}.
If we follow the pointers on the nodes, we obtain a "chain" of clusters, which either terminates at a terminal node, or
at a \defn{reciprocal nearest neighbor (\rnn) pair}, which is a pair of clusters that are each other's nearest
neighbor. 
The sequential NNC
algorithm~\cite{de1980classification, benzecri1982construction}
works by iteratively adding a node to a single chain through finding the nearest neighbor of
the \termi until an \rnn pair is found.
Each point is initially a singleton
cluster and a \termi of a single-node chain. The sequential algorithm picks an arbitrary node to start growing from.
After an \rnn pair is found, the \rnn pair is then merged, and the chain is grown again
to find another \rnn pair to merge. 
After $n-1$ merges, the algorithm finishes, producing a
hierarchy of clusters.


\myparagraph{Example} 
We now give an example of the definitions above by briefly describing running our \framework framework for parallel HAC on the small data set in \cref{fig:nnchain}.
This example uses the complete
linkage metric, where the distance between two clusters is the
distance of the farthest pair of points, one from each cluster.

Our framework
is based on the key insight that all
\rnn pairs can be merged simultaneously, which provides parallelism.
On each round, it
merges all \rnn pairs in parallel
(breaking
ties lexicographically\footnote{We use the ID of  the lexicographically first point in each cluster as the cluster's ID.} to prevent cycles). Before the first round, each point in $\{a,\dots,
f\}$ is represented by a chain with only one node,
and all points are singleton \termis.
The \rnn pairs are found by finding the nearest
neighbors of all \termis, which by definition are the clusters whose nearest
neighbors are unknown at the beginning of a round.
On the first round, we find the nearest neighbors
for all points in parallel. Now we have two chains, $\{f,e,d,c\}$ and
$\{a,b\}$. $\{e\}$, for example, is $\{f\}$'s successor.
$(a,b)$ and $(c,d)$ are two
 \rnn pairs, and so we merge them in parallel and create dendrogram nodes for clusters $\set{a,b}$ and $\set{c,d}$.  At the beginning of the
second round, $\set{a,b}$, $\set{c,d}$, and $e$ are \termis.  After we
find their nearest neighbors and grow the chain,  $(\set{a,b},e)$
is the only \rnn pair (we broke the tie  for $e$'s two nearest neighbors,
$\set{a,b}$ and $f$, by choosing $\set{a,b}$),
and so we merge it and create a dendrogram node for cluster $\set{a,b,e}$. We do not need to find the nearest
neighbor of $f$ in this round, because it is not a \termi and we know
that its nearest neighbor $e$ will not change (due to the reducibility property which will be defined
more formally in Section~\ref{sec:prelims}).
On the
third round, $\set{a,b,e}$, $\set{c,d}$, and $f$ are \termis. We find
the nearest neighbors for them and merge the \rnn pair $(\set{a,b,e},\set{c,d})$. Finally, on the fourth round, the \rnn pair
$(\set{a,b,c,d,e},f)$ is merged.

\vspace{5pt}

\framework achieves high
space efficiency and parallelism, which enables it to scale
HAC to large data sets that are orders of magnitude larger than those used in previous work.
There are two challenges in achieving 
both space efficiency and high parallelism. The first
challenge is to maintain all chains and merge reciprocal nearest
neighbor clusters correctly and efficiently \emph{in parallel}.
Unlike Jeon and Yoon's algorithm~\cite{jeon2014multi}, which is based on
locks (and has limited parallelism for large core counts), we use
lock-free approaches based on filtering and atomic operations
(\cref{sec:framework}).  The second challenge is to efficiently find
the nearest neighbors of clusters when growing the chain,
\emph{without} storing the distance matrix.
We introduce a range query optimization that significantly reduces the
number of distance computations used to find the nearest neighbor of
a cluster for low-dimensional data sets in Euclidean space (\cref{sec:nnfinding}), as well as a new caching technique
that stores a subset of previous distance computations that are likely
to be reused to further accelerate nearest neighbor searches
(\cref{sec:cache}). In the example in \cref{fig:nnchain},
the range query optimization avoids computing the distance between clusters
$\{e\}$ and $\{c,d\}$ in round 2 when $\{e\}$ searches for its nearest neighbor, because 
only clusters $\{a,b\}$ and $\{f\}$ will be within the range. 
The caching technique avoids storing all pairs of distances among the six 
points. In contrast, many previous methods~\cite{olson1995parallel,jeon2014multi,mullner2011modern, du2005novel,hung2014fast_protein_cluster, mullner2013fastcluster} require
a quadratic-space distance matrix and compute distances to all other clusters when searching for the nearest neighbor of a cluster.

We apply \framework to develop new linear-space
parallel HAC algorithms for the complete, Ward's, and average linkage
criteria.  Our framework can be applied for any linkage criteria that
satisfies the reducibility property, which ensures that the nearest
neighbor distance of clusters can never be smaller as clusters merge (defined
more formally in Section 2).

Though the worst case time complexity of our algorithms is $O(n^3)$, we observe that the running time is close to quadratic in practice on low-dimensional data sets because the range query is able to filter out many clusters. Many spatial, sensor, and computer vision data sets, where HAC is applicable, are low dimensional. 
In \cref{sec:exp},
we show experimentally on a variety of real-world and synthetic data sets (up to 16 dimensions) that our algorithms achieve 13.75--54.23x self-relative speedup on a
48-core machine with two-way hyper-threading. We also achieve 5.8--110.1x speedup over
the state-of-the-art parallel implementations. Our algorithms use up to 237.3x less space than existing implementations, and are able to scale to larger data sets with tens of millions of points, which existing algorithms are not able to handle.

We summarize our contributions below:
\begin{itemize}[topsep=1pt,itemsep=0pt,parsep=0pt,leftmargin=10pt]
    \item The \framework framework for parallel HAC using linear space.
    \item A range query optimization for fast nearest neighbor search for the complete, Ward's, and average  linkage criteria.
    \item A cache table optimization for reducing the number of cluster distance computations.
    \item Experiments showing that the algorithms in \framework achieve significant speedups over state-of-the-art.
\end{itemize}

\ifarxiv
Our source code is available at \url{https://github.com/yushangdi/parChain}.
\fi

%% file: background.tex
\section{Background}\label{sec:prelims}

The input to the \defn{hierarchical agglomerative clustering (HAC)} problem is a data set to be clustered and a linkage criteria that
specifies how distances between clusters are computed.
The output of HAC is a tree called a
\defn{dendrogram}, where the height of each dendrogram node represents the
dissimilarity between the merged two clusters according to the desired
linkage criteria.
A flat clustering,
which assigns the same ID to every object in the same cluster and different IDs to objects in different clusters, 
can be obtained by cutting the dendrogram at
some height. Thus, cutting the dendrogram at different heights gives
clusterings at different scales. An example of a dendrogram is shown
in Figure~\ref{fig:nnchain}(b).
In the rest of the section, we present our notations,
the three linkage criteria considered in this paper,
 and some relevant techniques used by our algorithm.

\myparagraph{Notation}
Let $v$ be a length-$d$ vector in $d$-dimensional space,
and
let $\|v\|$ denote the $L_2$ norm of $v$, i.e., $\|v\|=\sqrt{\sum_{i=1}^{d}|v[i]|^2}$ where $v[i]$ is the $i$'th coordinate of $v$.
$\bar{x}_A$ denotes the \defn{centroid} of cluster $A$, i.e., $\bar{x}_A = \frac{1}{|A|}\sum_{x \in A} x$, where the $x$'s are points in cluster $A$.
$\Var(A)$ denotes the \defn{variance} of cluster $A$, where $\Var(A) = \sum_{x \in A} \|x-\bar{x}_A \|^2$.
$\cld(A,B)$ denotes the \defn{distance} between clusters $A$ and $B$, and its formula depends on the linkage criteria.

\begin{table*}[t]
  \caption{Definitions, work, radius value, and optimizations used in our range query for different linkage criteria. 
  }\label{tbl:criteria}
  \begin{center}
  \begin{small}
      \begin{tabular}{ p{0.08\linewidth}  p{0.4\linewidth}  p{0.08\linewidth}  p{0.1\linewidth} p{0.3 \linewidth}}
      \toprule
      Linkage &  Cluster Distance $\cld(A,B)$ & Work & Radius & Optimizations\\
      \hline
      comp  & $\max_{x \in A, x' \in B} \|x-x'\|$ & $O(n^2)$ & $\beta$  & build $k$d-tree on all points\\
      Ward & $\sqrt{2(\Var(A \cup B) - \Var(A) - \Var(B))}$ $ = \sqrt{\frac{2|A||B|}{|A|+|B|} \|\bar{x}_A- \bar{x}_B\|^2}$ & $O(1)$  & $\beta\sqrt{\frac{|\cl_i|+n_{\min}}{2n_{\min} |\cl_i|}}$  & maintain cluster centroids and sizes \\ 
      avg-1 & $\frac{1}{|A||B|}\sum_{x \in A}\sum_{x' \in B} \|x-x'\|$ & $O(n^2)$& $\beta$  & -- \\ 
      avg-2 & $\frac{1}{|A| |B|}\sum_{x \in A}\sum_{x' \in B} \|x-x'\|^2  $ $= \|\bar{x}_A - \bar{x}_B\|^2 + \frac{\Var(A)}{|A|} + \frac{\Var(B)}{|B|}$ & $O(1)$&$\sqrt{\beta}$  & maintain cluster centroids, variances, and sizes \\
      \bottomrule
      \end{tabular}
  \end{small}
  \end{center}
  \end{table*}

\subsection{Linkage Criteria}\label{sec:prelims:funcs}
We now formally define the linkage criteria considered in this paper.
We use the Euclidean distance metric for all linkage criteria.
For average linkage, we also
consider the squared Euclidean distance metric.
The definitions of cluster distance under each linkage criteria and distance metric
are included in \cref{tbl:criteria}. We also include the work of each distance computation, the radius, and criteria-specific optimizations used in our range query optimization for different linkage criteria. The radius and optimizations will be discussed in more detail in \cref{sec:nnfinding}. We use \textit{comp, Ward, avg-1, and avg-2} to refer to complete linkage, Ward's linkage, average linkage with Euclidean distance metric, and average linkage with squared Euclidean distance metric, respectively.

\myparagraph{Complete Linkage}
In complete linkage~\cite{olson1995parallel, mullner2011modern},
the distance between two clusters is the maximum distance between a pair of points, one from each cluster.

\myparagraph{Ward's Linkage}
In Ward's linkage~\cite{ward1963hierarchical,murtagh2011ward}, or minimum variance linkage, the distance between two clusters is the increase in total 
variance if the two clusters merge.

\myparagraph{Unweighted Average Linkage}
In unweighted average linkage~\cite{sokal1958statistical, leibe2006efficient}, the distance between two clusters is the average distance between pairs of points, one from each cluster.

For Ward's linkage and average linkage with the squared Euclidean distance metric, we can be more space-efficient and compute the distance between two clusters in constant time by storing the mean and variance of every cluster, which takes only linear space overall. The newly merged cluster's mean and variance can be computed in constant time, where the new cluster's mean is an average of the means of the two original clusters, weighted by their sizes, i.e., $\bar{x}_{A\cup B} = \frac{|A|}{|A|+|B|}\bar{x}_A + \frac{|B|}{|A|+|B|}\bar{x}_B$. The variance is:
\begin{align*}
\Var(A\cup B) &= \frac{|A| \|\bar{x}_A - \bar{x}_{A\cup B}\|^2 + |B| \|\bar{x}_B - \bar{x}_{A\cup B}\|^2 }{|A|+|B|} \\
& +  \frac{ |A|\Var(A) + |B|\Var(B)}{|A|+|B|}
\end{align*}

\myparagraph{Lance-Williams Formula}
Many clustering metrics can be described using the Lance-Williams formula~\cite{lance1967general}. Given the distance between three clusters $A$, $B$, and $C$, we can obtain the distance between $A \cup B$ and $C$ using the following formula, with the coefficients for the metrics described above given in \cref{tbl:lance}:
$$ \cld(A\cup B, C)  = a_1\cld(A, C) +a_2\cld( B, C) +b\cld(A, B) +c\left|\cld(A, C) -\cld( B, C)\right|$$

The Lance-Williams formula allows for constant time distance
computation if we have the distances among clusters $A$, $B$, and $C$.
However, maintaining all these distances requires a distance matrix that
takes quadratic space.

\myparagraph{Reducibility}
We say a metric has the \defn{reducibility property}~\cite{bruynooghe1977methodes, murtagh1983survey, murtagh2017algorithms, jeon2014multi} if we have 
$\cld(A, B) < \cld(A \cup B, C)$ when $\cld(A, B)< \cld(A,C)$ or $\cld(A, B) < \cld(B, C)$.
All of the metrics introduced above satisfy the reducibility property. The reducibility property ensures that
the nearest neighbor of a cluster does not change unless one of the clusters merged is its nearest neighbor. For metrics that satisfy the reducibility
property~\cite{murtagh1983survey}, we can perform clustering using the
\textit{nearest-neighbor chain algorithm}~\cite{de1980classification,
  benzecri1982construction, juan1982programme, murtagh1983survey,
  gronau2007optimal, jeon2014multi} introduced in \cref{sec:intro}.
The reducibility property provides the parallelism in the nearest-neighbor chain algorithms since we can merge multiple reciprocal pairs simultaneously.

\begin{table}[t]
\caption{Coefficients for the Lance-Williams Formula~\cite{lance1967general}.}\label{tbl:lance}
\begin{center}
\begin{small}
\begin{tabular}{ccccc}
\toprule
& $a_1$ & $a_2$ & $b$ & \phantom{abc}$c$ \\ \hline
Complete linkage \phantom{abc} & $\frac{1}{2}$ & $\frac{1}{2}$ & 0 & \phantom{abc}$\frac{1}{2}$ \\ 
Ward's linkage \phantom{abc}& $\frac{|A|+|C|}{|A|+|B|+|C|}$ & $\frac{|B|+|C|}{|A|+|B|+|C|} $& $\frac{-|C|}{|A|+|B|+|C|}$& \phantom{abc}0\\
Average linkage\phantom{abc} & $\frac{|A|}{|A|+|B|}$& $\frac{|B|}{|A|+|B|}$ & 0 &\phantom{abc} 0 \\ 
\bottomrule
\end{tabular}
\end{small}
\end{center}
\end{table}

\subsection{Relevant Techniques}

\myparagraph{\kdt{}}
A \defn{\kdt{}} is a binary spatial tree where each internal node
contains a splitting hyperplane that partitions the points contained
in the node between its two children.
The root node contains all of the points, and the \kdt is constructed by recursing on each of its two children after splitting, until a leaf node is reached.
A leaf node contains at most $c$ points for a predetermined
constant $c$. The \kdt can be constructed in parallel by performing
the split and constructing each child in parallel.
The \textit{bounding box} of a node is the smallest rectangular
box that encloses all of its points.

\myparagraph{Nearest-Neighbor Query} A \defn{nearest-neighbor
  query} takes a set of points $\mathcal{P}$ and a query point $q$,
and returns for $q$ its nearest neighbor in $\mathcal{P}$ (besides itself if $q \in \mathcal{P}$).
An \defn{all-nearest-neighbor
  query} takes a set of points $\mathcal{P}$,
and returns for all points in $\mathcal{P}$ its nearest neighbor in $\mathcal{P}$ besides itself. The all-nearest-neighbor query can be performed efficiently using a dual-tree traversal~\cite{curtin2013treeind,march2010fast,curtin2015dualtreeruntime}, which we have parallelized.

\myparagraph{Range Query}
A \defn{range query} takes a set of points $\mathcal{P}$,
constructs a data structure to store the points, and reports or counts all points in some range $B$.
In this paper, we use balls to represent the ranges,  and we use  \kdt{s} to store the points.

\myparagraph{Other Parallel Primitives}
A parallel \defn{filter} takes an array $A$ and a
predicate function $f$, and returns a new array containing $a \in A$
for which $f(a)$ is true, in the same order that they appear in $A$.
A parallel \defn{reduce} takes as input a sequence $[a_1, a_2, \ldots , a_{n}]$ and an
associative binary operator $\oplus$, and returns
the overall sum (using $\oplus$) of the
elements $ (a_1 \oplus a_2
  \oplus \ldots \oplus a_{n})$.  
A parallel \defn{hash table} stores key-value pairs, and supports concurrent
insertions, updates, and finds.
\textsc{WriteMin} is a priority concurrent write that takes as input two
arguments, where the first argument is the location to write to and the
second argument is the value to write; on concurrent writes,
the smallest value is written~\cite{ShunBFG2013}. \textsc{WriteMax} is similar but writes the largest value.

%% file: framework.tex
\section{The \framework Framework}\label{sec:framework}
In this section, we 
present our framework \framework for parallelizing the
nearest-neighbor chain (NNC) algorithm, which works for all linkage criteria that satisfy the reducibility property explained in \cref{sec:prelims:funcs}. The
NNC algorithm exposes more parallelism than the
naive generic algorithm, where only the \rnn pair with minimum distance is merged, 
by allowing multiple \rnn pairs to be merged simultaneously.
Hence, our framework grows multiple chains and merges all \rnn pairs simultaneously in parallel. 

Jeon and Yoon's algorithm~\cite{jeon2014multi} uses a similar approach for to grow multiple chains in parallel, but it does merges \rnn pairs asynchronously.
It designates some threads for updating chains, and other threads for updating the cluster distances. 
Their algorithm also uses locks and requires quadratic memory for maintaining the distance matrix.
In contrast,
our algorithm proceeds in rounds where on each round, all chains are grown and all \rnn pairs are merged. Our algorithm is lock-free,
and uses linear space as we avoid using the distance matrix.
In addition, Jeon and Yoon's algorithm searches for the
nearest cluster naively by computing the distances to all other
clusters, whereas
we  have  optimizations for finding the nearest clusters
when growing the chain, which will be discussed in Sections~\ref{sec:nnfinding} and~\ref{sec:cache}.

\input{pseudo_fw}

\subsection{\framework{} Framework}
We now formally describe the \framework framework (Algorithm~\ref{alg:framework}). \framework  gives
rise to fast and space-efficient HAC algorithms.
The main idea of \framework is to avoid storing most cluster distances, and compute them on the fly using an optimized range search that considers only a small number of neighboring clusters.
We also cache some of the cluster distances to reduce computational cost.

The input to the algorithm is a set of $n$ points $\mathcal{P}$, a structure
$\distcomp$ that is used to compute the distances between clusters based on the linkage criteria,
and an integer $s \geq 0$ for the cache size. We store (cache) only $O(ns)$ cluster distances for an integer
$s \geq 0$ chosen at run time. The highlighted parts of
Algorithm~\ref{alg:framework} (Lines~\ref{alg:framework:cache-init}
and~\ref{alg:framework:cache-update-start}--\ref{alg:framework:cache})
are only required for $s > 0$, and we will discuss them in \cref{sec:cache}.
$\distcomp$ is able to compute the
distance between two clusters, and may
maintain some extra data to accelerate
 distance computations, such as the means and variances of
clusters.  It also specifies the Lance-Williams formula if $s > 0$,
which will be used for updating the entries of cached distances
between clusters.

\myparagraph{Initialization}
On Lines~\ref{alg:framework:initstart}--\ref{alg:framework:buildtree}, the algorithm initializes the required data structures.
It first creates $n$ dendrogram nodes (Line~\ref{alg:framework:initstart}) and $\chain$, a set of $n$ chain
nodes (Line~\ref{alg:framework:chain-init}). These nodes  are used for the singleton clusters at the beginning. We create a set of active clusters $\activeclusters$, initialized to contain all of the singleton clusters (Line~\ref{alg:framework:active-init}).
We also create a parallel hash table for each cluster to cache cluster distances if $s>0$ (Line~\ref{alg:framework:cache-init}).
In each chain node, we store its successor (succ), its predecessor (pred), and the distance to its predecessor (pred.$d$) if there is one. All chain nodes initially do not have any successor or predecessor.
 $\terminal$ represents the set of terminal nodes at the beginning of
 the round, and is initialized to contain the $n$ singleton chains (Line~\ref{alg:framework:termi-init}).
The algorithm also initializes a \kdt on the points
$P$. 
The \kdt (Line~\ref{alg:framework:buildtree}) is used to accelerate nearest cluster searches.

\myparagraph{Chain Growing and Merging}
On Lines~\ref{alg:framework:fnn}--\ref{alg:framework:merge}, in
parallel we grow all of the chains using the information in $\terminal$. We merge a node in $\terminal$ with its nearest neighboring cluster if they form an \rnn pair.
Specifically, on
Line~\ref{alg:framework:fnn}, to grow the chains we find the nearest neighbors of all
current \termis in $\terminal$ using a \kdt range search optimization, which will be
described in Section~\ref{sec:nnfinding}. \framework can quickly compute the distances
of a cluster to other clusters by considering only a small number of candidates, without
needing to maintain a distance matrix.
The nearest neighbors and the distances are stored in $\edges$.
On
Lines~\ref{alg:framework:pfor}--\ref{alg:framework:chain2}, we
update the successor and predecessor of each \termi in parallel to maintain the chains.
If a
\termi is the nearest neighbor of multiple clusters, the
\textsc{WriteMin} ensures that its predecessor is the cluster closest
to it.
 Then on
Line~\ref{alg:framework:findrnn}, we find all \rnn pairs
 using a parallel filter by checking for each \termi if its
successor has a successor that is itself.
All \rnn pairs are stored in an array
$\mathcal{M}$ with their distances. 
On Line~\ref{alg:framework:merge}, we create a new dendrogram node
$C_{i,j,\text{new}}$ to represent the merged cluster for each \rnn pair in
$\mathcal{M}$, which will have $\cl_i$ and $\cl_j$ as children, and store the distance between the merged clusters.

After the merges, we need to update the other data structures to prepare for next round.
On Lines~\ref{alg:framework:cache-update-start}--\ref{alg:framework:cache},
we update the cache tables with new distances 
Section~\ref{sec:cache}. On Line~\ref{alg:framework:dist}, 
we update the extra data structures, such as the \kdt and clusters' mean and variance.
On Line~\ref{alg:framework:update-active}, we update the set of active clusters by including active clusters not in $\mathcal{M}$ (not merged this round) and the newly merged clusters from this round.
Finally, on
Line~\ref{alg:framework:updateterm}, we
obtain the new set of \termis $\terminal$ using a parallel filter on the active clusters $\activeclusters$.

\myparagraph{Work and Space Complexity} We summarize the time and space complexity of the state-of-the-art algorithms in \cref{tbl:complexity}. \framework is the only algorithm that requires linear memory and works for a broad set of linkage criteria (any that satisfy the reducibility property). The main computational cost in our
framework is in finding the nearest neighbors of the \termis on each
round, and updating the cache tables and other data structures
maintained by the distance structure $\distcomp$.
Sections~\ref{sec:nnfinding} and~\ref{sec:cache} present our novel
approaches for efficiently computing nearest neighbors efficiently
with low space. 

We now analyze the work of our framework. Let $\terminal_i$ and $\activeclusters_i$ be the sets $\terminal$ and $\activeclusters$ at the beginning of round $i$, respectively. The initialization (Lines~\ref{alg:framework:initstart}--\ref{alg:framework:buildtree}) take $O(n \log n)$ work, dominated by the \kdt construction work on Line~\ref{alg:framework:buildtree}. Lines~\ref{alg:framework:pfor}--\ref{alg:framework:cache} and Lines~\ref{alg:framework:update-active}--\ref{alg:framework:updateterm} take $O(\sum |\terminal_i|)$ work across all rounds, plus the cost of all cluster distance computations, denoted by $D$. Line~\ref{alg:framework:dist} takes $O(\sum |\activeclusters_i| \log |\activeclusters_i|)$ work because we need to re-construct the \kdt of cluster centroids in this step. Finally, Line~\ref{alg:framework:fnn} takes $O(\sum |\activeclusters_i||\terminal_i|)$ work because for each terminal node, we need a range query on the \kdt of cluster centroids. Thus, we have that the work of \framework is $O(D+M)$, where $M=\sum |\activeclusters_i|(|\terminal_i| + \log |\activeclusters_i|)$. 
In the worst case, the work is $O(n^3)$, but 
we show in
Section~\ref{sec:exp} that in practice both $M$ and $D$ are close to quadratic and \framework is orders of
magnitude faster than the $O(n^2)$ work
algorithms~\cite{mullner2013fastcluster}, even using a
single thread. 
We expect our algorithm to give improvements on most low-dimensional data sets.

The space usage of our framework is $O(n(1+s))$ because all data
structures except the caches require linear memory, and the caches
require $O(ns)$ memory. The \kdt requires memory linear
in the number of points in the tree.

\begin{table}[t]
    \caption{Worst-case work and space bounds of state-of-the-art HAC algorithms. 
  $^*$The authors of \cite{nguyen2014sparsehc,loewenstein2008efficient, sun2009esprit}  do not report the work complexity.
  }\label{tbl:complexity}
  \begin{center}
  \begin{small}
      \begin{tabular}{ c c c c}
      \toprule
      Algorithm & Work & Space & Restrictions\\
      \hline
      ParChain & $O(n^3)$ & $O(n)$ & Reducibility\\ 
      NN-Chain ~\cite{jeon2014multi, mullner2013fastcluster, Virtanen2020scipy} & $O(n^2)$ & $O(n^2)$  & Reducibility\\ 
      Generic ~\cite{olson1995parallel, mullner2013fastcluster, scikit-learn, hung2014fast_protein_cluster} & $O(n^2 \log n)$ & $O(n^2)$  & Lance-Williams\\ 
      Althaus et al. \cite{althaus2014greedy} & $O(n^3)$ & $O(n)$ & Complete Linkage\\ 
      Batch Processing~\cite{nguyen2014sparsehc,loewenstein2008efficient, sun2009esprit} & $^*$ & $O(n^2)$ & Disk-based\\ 
      \bottomrule
      \end{tabular}
  \end{small}
  \end{center}
  \end{table}

%% file: pseudo_fw.tex
\begin{algorithm}[!t]
\DontPrintSemicolon
\fontsize{8pt}{8pt}\selectfont
\KwData{$n$ points $\mathcal{P}$, distance structure $\distcomp$, and cache size $s$}
\KwResult{Dendrogram tree $\hierarchy$}
Initialize $n$ dendrogram leaf nodes $\cl_0,\ldots,\cl_{n-1}$ to each represent a singleton cluster (a point).\; \label{alg:framework:initstart}
Initialize $\chain$, a set of $n$ chain nodes, where each $\chain_i$ represents a singleton cluster. \; \label{alg:framework:chain-init}
Initialize $\activeclusters = \set{C_0,\ldots,C_{n-1}}$, the set of active clusters. \; \label{alg:framework:active-init}
\custo{Create cache tables $\set{\cache_i}$ for clusters, each of size $s$.}\;\label{alg:framework:cache-init}
Terminal nodes $\terminal$ = $\set{\chain_0, \dots, \chain_{n-1}}$.\; \label{alg:framework:termi-init}
$\tree$ = \kdt $\tree_\mathcal{P}$\;\label{alg:framework:buildtree}
\While{$|\activeclusters|>1$}{
  $\edges$ = $\mathtt{find\_nearest\_neighbors}$($\tree$, $\distcomp$, $\chain$, $\terminal$)\; \label{alg:framework:fnn}
  \tcp*[h]{Below, $C_i$ is the \termi and $C_j$ is its nearest neighbor.}\;
  \PFor{$ (\cl_i, \cl_j, d) \in \edges$\label{alg:framework:pfor}}{
    $\chain_i$.succ = $j$\;\label{alg:framework:chain1}
    $\writemin(\chain_i$.pred, $(j, d)$)\tcp*{The pair with the minimum  $d$  is written.}\label{alg:framework:chain2}
  }
  $\mathcal{M}$ = parallel\_filter($\edges$, is\_R-NN())\;\label{alg:framework:findrnn}
\PFor{$(\cl_i, \cl_j, d) \in \mathcal{M}$}{
  $C_{i,j,\text{new}}$ = merge($C_i$, $C_j$, $d$)\;\label{alg:framework:merge}
  \custo{\If{$s > 0$}{\label{alg:framework:cache-update-start}
    \PFor{$(\cl_i, \cl_j, d) \in \mathcal{M}$}{
    $\mathtt{update\_cached\_dists}$($\cl_i, \cl_j, d$)\;\label{alg:framework:cache}
    }
    }
  }
  }
$\distcomp$.update($\tree$, $\mathcal{M}$)\; \label{alg:framework:dist}
$\activeclusters$ = parallel\_filter($\activeclusters$, not\_in\_$\mathcal{M}$()) $\ \bigcup\ $
$\{C_{i,j,\text{new}} \ \mid\ (C_i,C_j,d)\in \mathcal{M} \}$
  \; \label{alg:framework:update-active}
$\terminal$ = parallel\_filter($\activeclusters$, is\_terminal())\; \label{alg:framework:updateterm}
} 
\Return dendrogram root node\;
\caption{\framework Framework}\label{alg:framework}
\end{algorithm}

%% file: nn_finding.tex
\section{Nearest-Neighbor Finding}\label{sec:nnfinding}
We will now describe how to efficiently perform nearest-neighbor
finding (Line~\ref{alg:framework:fnn} of
Algorithm~\ref{alg:framework}) for the three linkage
criteria: complete, Ward's, and average linkage. We assume that we
compute distances between clusters on the fly.
We describe an optimization in Section~\ref{sec:cache} that uses cache tables to store some of the distances.

While a standard nearest-neighbor search is done on points, we are searching for nearest neighbors of clusters with distances based on the linkage criteria.
Our \kdt{s} store centroids of \emph{clusters of points}, which we use to find nearby clusters to our query cluster. We then perform exact distance computations from our query cluster to these clusters.
Unlike in standard nearest-neighbor searches, it is harder to prune in our case as the distances between clusters centroids do not necessarily correspond to distances between clusters.
Instead, we compute a different search area for each cluster
based on an upper bound on the distance between the query
cluster and its nearest neighbor. This upper bound can be a distance
between the query cluster and any other cluster. We provide a novel
heuristic for finding a good upper bound on the distance to the
nearest cluster, and only search within this distance in Sections~\ref{sec:ward}--\ref{sec:complete}. In Figure~\ref{plot:norange}, we
present the performance of using our optimized range query compared to
the naive method of computing the distances to all other clusters to
find the nearest neighbor. We see that our optimized range query gives
a 7.8--1892.4x speedup on the two example data sets.

\input{fig_exp_norange}
\input{pseudocode_range}

\myparagraph{Algorithm}
Given the \kdt $\tree$ built on the centroid of clusters, distance structure $\distcomp$, 
chain nodes $\chain$,
and set of \termis
$\terminal$,
Algorithm~\ref{alg:nn} finds the nearest cluster and the distance to
it for each terminal node's cluster and stores them in $\edges$.

When finding the nearest neighbor of cluster $\cl_i$,
we search all points within some ball $Ball(\bar{x}_{\cl_i}, r)$  obtained from $getBall_\distcomp(i,\beta)$, which is a ball centered at centroid $\bar{x}_{\cl_i}$ with radius $r$. The radius $r$ 
depends on the linkage method of $\distcomp$ and an distance $\beta$
between $\cl_i$ and another cluster (Lines~\ref{alg:range:pfor}--\ref{alg:range:rq} of Algorithm~\ref{alg:nn}).  
Since we use centroid distances, $\distcomp$ rebuilds $\tree$ to be a \kdt of only the
centroids of current clusters at the end of each round of Algorithm~\ref{alg:framework}
(Line~\ref{alg:framework:dist} of Algorithm~\ref{alg:framework}). 
If $\cl_i$
has a predecessor, we set the distance $\beta$ to be the distance between $\cl_i$ and its
predecessor (Lines~\ref{alg:nn:if}--\ref{alg:nn:pr} of Algorithm~\ref{alg:nn}). Otherwise, we
find the distance to another cluster for computing the radius for the search.
Specifically, we use the distance to the cluster 
whose centroid is the closest to the current cluster, 
which can be computed using a parallelized nearest-neighbor query on the \kdt of centroids~\cite{callahan1993optimal}
(Lines~\ref{alg:range:else2}--\ref{alg:range:ar1} of Algorithm~\ref{alg:nn}).

For the range query on Line~\ref{alg:range:rq} of Algorithm~\ref{alg:nn}, we use the parallel range query
in Algorithm~\ref{alg:range}. Given a query cluster $\cl_i$, a \kdt node $Q$, a ball representing the range, a linkage function $\distcomp$, and a set $\edges$ of pairs of nearest neighbor candidates and distances of terminal nodes, the algorithm processes all of $Q$'s points that are in the ball to update the nearest neighbor candidates in $\edges$.
Since we only process the points in the ball, on Line~\ref{alg:range:overlap}, the range query terminates if the bounding box of the tree node does not overlap with the query range.
Otherwise, the range query will either process all of the points both in the node and in the ball using the \textit{update\_nearest\_neighbor} subroutine (Algorithm~\ref{alg:updatenn}) if it is a leaf node (Lines~\ref{alg:range:if}--\ref{alg:range:leaf} of Algorithm~\ref{alg:range})
or recurse on its two children in parallel (Lines~\ref{alg:rangequery:else}--\ref{alg:range:spawn2} of Algorithm~\ref{alg:range}).

In each update\_nearest\_neighbor($\cl_i, \cl_j$) call,
we check if some cluster $\cl_j$ is closer to $\cl_i$ than its current nearest neighbor candidate, and if so we update $\cl_i$'s nearest neighbor in $\edges$ with Algorithm~\ref{alg:updatenn}. We also update $\cl_j$'s nearest neighbor to be $\cl_i$ if $\cl_i$ is closer to $\cl_j$ than its current nearest neighbor candidate. 
In Algorithm~\ref{alg:updatenn}, if $s=0$, we will compute the distance between $\cl_i$ and $\cl_j$ on the fly (Lines~\ref{alg:updatenn:compute}, \ref{alg:updatenn:update2}, and \ref{alg:updatenn:update3}). If
$s>0$, we will first check the cache and use a cached distance if
possible (we describe more details  in Section~\ref{sec:cache}).

As an optimization for the first round, we know that the distances between  clusters is
exactly the same as the distances between their centroids in the first round, and thus we can efficiently prune searches in the \kdt{s}.
Therefore, we use an all-nearest-neighbor query for the first round, which we implemented by parallelizing the dual-tree traversal algorithm by March et al.~\cite{march2010fast}. At a high level, our algorithm processes recursive calls of the dual-tree traversal in parallel and uses \writemin to update the nearest neighbors of points.
A dual-tree traversal allows more pruning than 
when running individual nearest neighbor queries for each point.

In the rest of the section, we will
describe
the radius of the search ball for each linkage method. We will show that a cluster's nearest neighbor must have its centroid inside the ball. 

\input{fig_range}

\input{ward_avg}

\input{comp}

%% file: fig_exp_norange.tex
\begin{figure}[t]
\begin{center}
\includegraphics[width=\columnwidth]{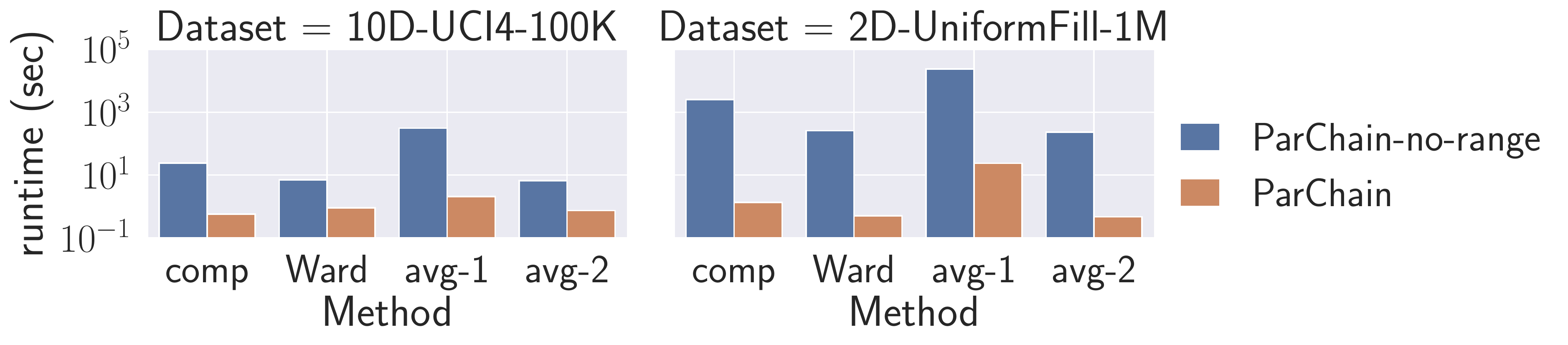}
\caption{Runtimes on 48 cores with two-way hyper-threading of using our optimized range
query compared to not using the range query and computing the distances to all
other clusters on the fly to find the nearest neighbor. }
\label{plot:norange}
\end{center}
\end{figure}

%% file: pseudocode_range.tex
\begin{algorithm}[!t]
\DontPrintSemicolon
\fontsize{8pt}{8pt}\selectfont
\KwData{\kdt $\tree$, distance structure $\distcomp$, chains $\chain$, and set of terminal nodes
$\terminal$}
\KwResult{nearest neighbors of nodes in $\terminal$}
Initialize $\edges$ with a (Null, $\infty$) entry for each \termi. \;
\PFor{$\cl_i \in \terminal$ \label{alg:range:pfor}}{
\tcc{$\chain_i$ is the chain node of $\cl_i$}
\eIf{$\chain_i$.pred $\neq$ Null
}{ \label{alg:nn:if}
$\beta = \chain_i$.pred.$d$\;\label{alg:nn:pr}
}{ \label{alg:range:else2}
$\beta = $ distance to a nearby cluster \;\label{alg:range:ar1}
}
\tcp*[h]{range query updates $\edges$}\;
$\mathtt{range\_query}(\cl_i, \tree, getBall_\distcomp(i, \beta), \distcomp, \edges)$   \label{alg:range:rq}
}
\Return $\edges$\;
\caption{Find Nearest Neighbor}\label{alg:nn}
\end{algorithm}

\begin{algorithm}[!t]
\DontPrintSemicolon
\fontsize{8pt}{8pt}\selectfont
\KwData{query node $\cl_i$, \kdt node $Q$, $Ball$, $\distcomp$, $\edges$}

\lIf{$Q$ does not overlap with $Ball$}
\Return
\label{alg:range:overlap}

\eIf{\upshape $Q$ is a leaf node\label{alg:range:if}}{
\lFor{\upshape $\bar{x}_{\cl_j} \in Q$ and $\bar{x}_{\cl_j} \in$ Ball}{
update\_nearest\_neighbor($\cl_i$, $\cl_j$, $\edges$, $\distcomp$)\label{alg:range:leaf}
}
}{\label{alg:rangequery:else}
\textbf{par\_do} (RangeQuery($\cl_i$, $Q$.left, $Ball$, $\distcomp, \edges$),\; \hspace{27pt} RangeQuery($\cl_i$, $Q$.right, $Ball$, $\distcomp$, $\edges$))\;\label{alg:range:spawn2}
}

\caption{RangeQuery}\label{alg:range}
\end{algorithm}

\begin{algorithm}[!t]
\DontPrintSemicolon
\fontsize{8pt}{8pt}\selectfont
\KwData{cluster $\cl_i$, cluster $\cl_j$, distance structure $\distcomp$, cache tables $\cache_i$ and $\cache_j$, and set $\edges$}
\custo{\If{$s>0$}{
		d = get\_cached\_dist($i,j$)\;\label{alg:updatenn:query}
		\If{$d\neq$ NOT\_FOUND}{ \label{alg:updatenn:if}
		  \writemin($\edges[\cid_i], (\cid_j, d)$)\;
                  \writemin($\edges[\cid_j], (\cid_i, d)$)\; \label{alg:updatenn:update1}
		  \Return \label{alg:updatenn:return}
		}
}}

$d = \distcomp$.dist($C_i, C_j$)\; \label{alg:updatenn:compute}
\custo{\lIf{$s>0$}{insert $\entry{\cid_i}{i}{d}$ into $\cache_j$ and $\entry{\cid_j}{j}{d}$ into $\cache_i$}} \label{alg:updatenn:insert} 
$\writemin(\edges[\cid_i], (\cid_j, d))$\; \label{alg:updatenn:update2}
$\writemin(\edges[\cid_j], (\cid_i, d))$\; \label{alg:updatenn:update3}
\caption{Update Nearest Neighbor}\label{alg:updatenn}
\end{algorithm}

%% file: fig_range.tex
\begin{figure}[!t]
  \centering
\includegraphics[trim=0 0 0 0, clip,width=0.85\columnwidth]{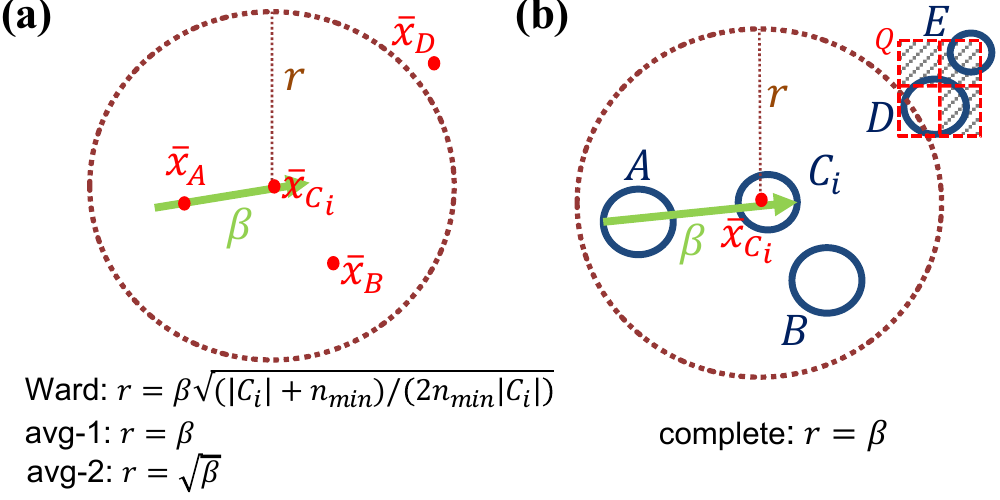}
\caption{Search area ball with radius $r$ in range queries for the linkage metrics. The blue circles are clusters. The red points are cluster centroids. The green arrows specify the predecessor $A$ of $C_i$, and the distance between them is $\beta$. The red boxes are the bounding boxes of a \kdt node $Q$ and its four descendants. 
In (a), $\cld(\cl_i,B)$ is computed because $\bar{x}_B$ is in the ball; however, $\cld(\cl_i,D)$ is not computed because $\bar{x}_D$ is outside the ball. $n_{\min}$ is the size of the smallest current cluster.
In (b), $\cld(\cl_i,B)$ is computed because $B$ is completely in the ball. $\cld(\cl_i,D)$ will not be computed, because the three shaded bounding boxes of \kdt nodes do not intersect with the ball, and so some of $D$'s points will not be included in the count. 
We only compute the distance to a cluster if all of its points  are included in the count. 
}
\label{fig:range}
\end{figure}

%% file: ward_avg.tex
\subsection{Ward's Linkage}\label{sec:ward}
In Ward's linkage, $\cld(\cl_i,B)_{\smb{Ward}} = \sqrt{\frac{2|\cl_i||B|}{|\cl_i|+|B|} ||\bar{x}_{\cl_i} - \bar{x}_B||^2}$.
For the range query, we can use a ball with radius $r = \beta\sqrt{\frac{|\cl_i|+n_{\min}}{2n_{\min} |\cl_i|}}$, where $\beta$ is the distance between $\cl_i$ and some cluster $A$ and $n_{\min}$ is the size of the smallest current cluster. We can obtain $n_{\min}$ using a parallel reduce on the sizes of all clusters. Figure~\ref{fig:range}(a) illustrates the range search for Ward's linkage.

Since $\beta = \cld_{\smb{Ward}}(\cl_i,A)$, any cluster $B$ that is closer to $\cl_i$ than $A$ must have
$\|\bar{x}_{\cl_i} - \bar{x}_B\|^2 \leq \beta^2\frac{|\cl_i|+|B|}{2|\cl_i||B|}$ .
The right-hand side of the inequality becomes smaller for larger $|B|$, thus we can upper bound the distance between $\cl_i$'s centroid and $B$'s centroid (i.e., $\|\bar{x}_{\cl_i} - \bar{x}_B\|$) by $r = \beta\sqrt{\frac{|\cl_i|+n_{\min}}{2n_{\min} |\cl_i|}}$.
Therefore, no cluster outside of the ball centered at $\bar{x}_{\cl_i}$ with radius $r$ can be closer to $\cl_i$ than $A$, and thus we only need to search for $\cl_i$'s nearest neighbor inside this ball.

\subsection{Average Linkage}\label{sec:average}
In average linkage, the distance between two clusters is the average
distance between all pairs of points, one from each cluster. For the range query, we use a
ball centered at $\bar{x}_{\cl_i}$ with radius $r=\beta$ and
$r=\sqrt{\beta}$ for Euclidean distance metric and squared Euclidean distance metric, respectively. As before, $\beta$ is the distance between $\cl_i$ and some cluster $A$.
\ifarxiv
Now, we show that the nearest neighbor $B$ of $\cl_i$ must have its centroid inside the ball. First, we show this for the Euclidean distance metric.
Observe that if we have
$\|\bar{x}_{\cl_i} - \bar{x}_B\| \leq \cld_{\smb{avg-}1}(\cl_i,B)$,
then we have $\|\bar{x}_{\cl_i} - \bar{x}_B\| \leq \cld_{\smb{avg-}1}(\cl_i,B) \leq \cld_{\smb{avg-}1}(\cl_i,A) = \beta = r$, and thus the centroid $\bar{x}_B$ is within the ball. We obtain
$\|\bar{x}_{\cl_i} - \bar{x}_B\| \leq \cld_{\smb{avg-}1}(\cl_i,B)$ as follows:

\begin{align*}
\cld_{\smb{avg-}1}(\cl_i,B)&=\frac{1}{|\cl_i||B|}\sum_{x_k \in \cl_i}\sum_{x_j \in B} \|x_k-x_j\| \\
&\geq \frac{1}{|\cl_i||B|}\bigg\|\sum_{x_k \in \cl_i}\sum_{x_j \in B} (x_k-x_j)\bigg\| \\
&=\frac{1}{|\cl_i||B|} \bigg\||B|\sum_{x_k \in \cl_i}x_k - |\cl_i|\sum_{x_j \in B} x_j \bigg\|  \\
&=\bigg\| \sum_{x_k \in \cl_i} \frac{x_k}{|\cl_i|} - \sum_{x_j \in B} \frac{x_j}{|B|} \bigg\| \\
&=\|\bar{x}_{\cl_i} - \bar{x}_B\|
\end{align*}
\else
The nearest neighbor $B$ of $\cl_i$ must have its centroid inside the ball, i.e., $\|\bar{x}_{\cl_i} - \bar{x}_B\| \leq \cld_{\smb{avg-}1}(\cl_i,B)$. 
The proof is provided in the full version of our paper~\cite{yu2021parchain}.
\fi

Similarly, for the squared Euclidean metric, we have
$\|\bar{x}_{\cl_i} - \bar{x}_B\|^2 \leq \cld_{\smb{avg-}2}(\cl_i,B)$,
which leads to $\|\bar{x}_{\cl_i} - \bar{x}_B\|^2 \leq \cld_{\smb{avg-}2}(\cl_i,B) \leq \cld_{\smb{avg-}2}(\cl_i,A) = \beta = r^2$.
$\|\bar{x}_{\cl_i} - \bar{x}_B\|^2 \leq \cld_{\smb{avg-}2}(\cl_i,B)$ holds since variances are non-negative and $\cld(\cl_i,B)_{\smb{avg-}2} = \|\bar{x}_{\cl_i} - \bar{x}_B\|^2 + \frac{\Var(\cl_i)}{|\cl_i|} + \frac{\Var(B)}{|B|}$.
Figure~\ref{fig:range}(a) illustrates the range search for average linkage with the Euclidean and squared Euclidean distance metrics.

%% file: comp.tex
\subsection{Complete Linkage}\label{sec:complete}
In complete linkage, the distance between two clusters is the maximum distance between a pair of points, one from each cluster. 
For the range query, we use a ball with radius $r=\beta$ centered at centroid $\bar{x}_{\cl_i}$, where $\beta$ is the distance between $\cl_i$ and some cluster. By definition of the complete linkage function, the cluster distance must be no smaller than distance between their centroids, and so the nearest neighbor of $\cl_i$ has its centroid within the search ball.

\myparagraph{Range Query Optimization}
For complete linkage, we can reduce the number of cluster distance computations by only computing the distance to a cluster if it is completely within the search ball. With this observation, we can optimize the algorithm by keeping
the \kdt to be $\tree_P$, the \kdt of all points, and avoiding updating it to be the \kdt of centroids on every round.
Figure~\ref{fig:range}(b) illustrates the optimized range search for complete linkage. We will prove the correctness of this optimization at the end of the subsection.

Since now $\tree$ is always $\tree_P$, we need to slightly modify Algorithm~\ref{alg:nn} and Algorithm~\ref{alg:range}. On  Line~\ref{alg:range:ar1} of Algorithm~\ref{alg:nn}, we search for the point $p \notin \cl_i$ closest to $\bar{x}_i$ in $\tree_P$, and let $\beta$ be the distance between $\cl_i$ and the cluster of this point. We can use a parallel union-find structure~\cite{Dhulipala2020} to ignore  points in $\cl_i$.
For Algorithm~\ref{alg:range}, the range query might be able to terminate before Line~\ref{alg:range:if} if the tree node satisfies some conditions.
For each range search, we keep a count that eventually upper bounds the number of points within the ball for each cluster. In each for-loop on Line~\ref{alg:range:leaf} of Algorithm~\ref{alg:range}, we now loop over points $p$ instead of centroids, and we atomically increment the count for $p$'s cluster by 1 because this means we have found one more point in this cluster that is within the ball.
The cluster IDs can also be maintained and queried using the parallel union-find data structure.
Right before Line~\ref{alg:range:if} of Algorithm~\ref{alg:range}, if all points in the \kdt node $Q$ are from the
same cluster $C$, we atomically increment the count of cluster $C$ by the size of the node and prune the search; otherwise, we continue the search and recurse on the children. This gives an upper bound on the number of points in the cluster within the ball, because the ball lies inside the \kdt bounding boxes traversed. 

We preprocess the tree such that in the range search we can determine in constant time if all points in the node are from the same cluster, and if so which cluster it is. Specifically, we mark the
\kdt nodes with a cluster ID if all points in the node are from the
same cluster, or with $\nocid$ if the points in the node belong to
multiple clusters.
This can be computed by recursively
checking the ID of the two children of a node starting from the root,
and storing the cluster ID of the children if all of their
points are from the same cluster. We update this information on every round.

After processing a point or a node, if we incremented the count of a cluster $\cl$, we check if the count of  $\cl$ is equal to the size of $\cl$. If so,
this means that all of $\cl$'s points may be within distance $r=\beta$. In this case,
we compute the distance between the $\cl_i$ and this cluster, and use a \writemin to update the nearest neighbor of $\cl_i$ in $\edges$ (Lines~\ref{alg:updatenn:compute}, \ref{alg:updatenn:update2}, and \ref{alg:updatenn:update3} of Algorithm~\ref{alg:updatenn}).

Finally, we show below that $\cl_i$'s nearest neighbor $B$ must be a cluster completely within search area by claiming that clusters with points outside the ball must have a distance larger than $r$ to $\cl_i$. Since $r$ is the distance between $\cl_i$ and some cluster, $B$ must have a distance no larger than $r$ to $\cl_i$.
Suppose the distance of the furthest point pair between $C_i$ and $B$ is $\cld_\smb{comp}(\cl_i, B) = d(p,q)$. Since the average Euclidean distance between points in two clusters is not smaller than the distance between their centroids
\ifarxiv
(shown in Section~\ref{sec:average}),
\else
(shown in the full paper~\cite{yu2021parchain}),
\fi
applying this property to $\cl_i$ and $\{q'\}$ for any point $q' \in B$, we see there must exists some $p' \in \cl_i$ such that $d(\bar{x}_{C_i}, q') \leq d(p', q')$.
Since $(p,q)$ is the furthest point pair, we have that
$d(\bar{x}_{C_i}, q') \leq d(p', q') \leq d(p,q)$. Thus, if $\cld_{\smb{comp}}(\cl_i, B) = d(p,q) \leq r$, then all points in $B$ must be within $Ball_{\smb{comp}}(\bar{x}_{\cl_i}, r)$.
As a result, we only need to consider a cluster as the nearest neighbor candidate of $\cl_i$ and actually compute the distance to it if all of its points are inside the ball.

\myparagraph{Dual-Tree Traversal}
When computing cluster distances (Line~\ref{alg:updatenn:compute} of \cref{alg:updatenn}) for complete linkage, we use our parallel dual-tree traversal algorithm described earlier in the section. We 
need to find the distance of the farthest pair of points, and so we use \textsc{WriteMax} instead of \textsc{WriteMin} for storing the farthest distance seen.
In order to perform the dual-tree traversals, $\distcomp$ creates a \kdt for each cluster at the end of each round (Line~\ref{alg:framework:dist} in Algorithm~\ref{alg:framework}).

%% file: cache.tex
\section{Caching Inter-Cluster Distances}\label{sec:cache}
For some linkage function and metric combinations, such as average
linkage with the Euclidean distance metric, computing inter-cluster
distances can be expensive. We can avoid some recomputations of
cluster distances by caching some previously computed distances for
each
cluster $\cl_i$ using a cache table $\cache_i$, represented using a
parallel hash table. Users can specify a constant size $s$ of each
cache based on the available memory.  The total memory usage is
$O(n(1+s))$, which is less than the quadratic memory required by the
distance matrix approaches. 
Sometimes, a larger table will lead to
faster computations because we can cache more distances and avoid more
recomputations. 
Due to the optimizations in \cref{sec:nnfinding}, the
distances that we compute will tend to be close to
$\cl_i$, and hence stored in $\cache_i$. These distances are more
likely to be reused in future nearest neighbor queries.

We present a comparison of running times of average linkage with the
Euclidean distance metric on several data sets using different cache
sizes in Figure~\ref{plot:cachesize}. We see that using caching
improves the running times by up to a factor of 8.98x compared to
not using caching. We found similar trends on other data sets.  We
will discuss more about our implementation's memory usage in
Section~\ref{sec:exp}.  

\input{fig_exp_cachesize}

In the rest of the section, we assume $s >
0$ and describe how to query cluster distances
from the cache tables, insert new entries after computing cluster
distances during nearest neighbor queries, and update the tables after
merging clusters.

\myparagraph{Querying and Inserting Distances between Clusters}
The cache tables can be used to reduce cluster distance computations
because we can insert the computed distances to the tables and query
for them if we want to use them again. 
Now we describe how the cache is used to update the nearest neighbor
candidate in the nearest neighbor search
(Algorithm~\ref{alg:updatenn}). With $s>0$, we might have already
cached the distance $\cld(\cl_i, \cl_j)$ in one or both of the tables
$\cache_i$ and $\cache_j$ when we find $\cl_j$ in $\cl_i$'s range.
Therefore, we first query for the distance in the cache tables (Line~\ref{alg:updatenn:query}), and
only compute the distance if the return value is NOT\_FOUND; otherwise we can
directly use the queried distance to update the nearest neighbor candidate in $\edges$ (Lines~\ref{alg:updatenn:if}--\ref{alg:updatenn:return}). 
If we compute
the distance (Line~\ref{alg:updatenn:compute}), we will attempt to insert it into both of the tables (Line~\ref{alg:updatenn:insert}). The
insertion may fail for a cache table if it is full, i.e., it already
contains $s$ entries.
Since we insert distances between $\cl_i$ and the clusters that are
within its search range in all rounds so far, the distances
stored in $\cache_i$ are likely to be between $\cl_i$ and nearby
clusters.  Thus in later rounds, these cached distances are more
likely to be queried.
On Lines~\ref{alg:updatenn:update2}--\ref{alg:updatenn:update3}, \writemin updates the nearest neighbors of $\cl_i$ and $\cl_j$ in $\edges$.

When querying
$\cld(\cl_i, \cl_j)$ with get\_cached\_dist($i,j$) (Line~\ref{alg:updatenn:query}), we search for the
entry with key $i$ in $\cache_j$, and the entry with key $j$
in $\cache_i$. If in a cache table, the key does not exist, then the query fails. 
If the queries in
both tables fail, we return NOT\_FOUND.  If the search is
successful in one of the tables, we return the distance stored in the
table.  We search in both cache tables since the caches are of limited
size, and so the distance could potentially be stored in just one of
the two tables.

\myparagraph{Updating Cache Tables after Merging Clusters}
We now describe how to update the entries in the cache tables after clusters are merged (Lines~\ref{alg:framework:cache-update-start}--\ref{alg:framework:cache} of \cref{alg:framework}). 
If during a round $\cl_i$ and $\cl_j$ are merged into a new cluster $\cl_k$, we will try to compute the distance between $\cl_k$ and all
clusters $\cl_{g}$ whose subclusters' distance(s) with $\cl_i$ or $\cl_j$ are stored in $\cache_i \cup \cache_j$. These distances  can be used to accelerate the computation of $\cld(\cl_k, \cl_{g})$ using the Lance-Williams formula described in \cref{sec:prelims:funcs}. 

\input{pseudo_cache}

\input{fig_cache}

The update\_cached\_dists function called on
\cref{alg:framework:cache} of \cref{alg:framework} is presented in
\cref{alg:updatecache}.
On Line~\ref{alg:updatecache:loop}, we loop
over the distances $d'$ in the cache tables of $\cache_i$ and $\cache_j$.
Without
loss of generality, assume $d' = \cld(\cl_i, \cl_\ell)$ is a distance in $\cache_i$
between $\cl$ and some cluster $\cl_\ell$ (the case
for an entry in $\cache_j$ is similar).
\cref{alg:updatecache:continue} skips over the entries that represent distances between $C_i$ and $C_j$, since they are now merged.
Otherwise, there are two cases.
In case (1),  $\cl_\ell$ is also a
cluster merged in this round
(Lines~\ref{alg:updatecache:1}--\ref{alg:updatecache:1d}), and we let $\cl_g$
be the cluster that $\cl_\ell$ merged into. We compute the new
distance $\cld(\cl_k, \cl_{g})$ on Line~\ref{alg:updatecache:1d} and
insert the new distance into the caches of both clusters $\cl_k$ and
$\cl_{g}$ on Line~\ref{alg:updatecache:insert}. In case (2),
$\cl_\ell$ is not a new cluster merged in this round
(Lines~\ref{alg:updatecache:2}--\ref{alg:updatecache:2d}), and we have
 $\cl_g=\cl_\ell$.
We can also
use $d'$ to accelerate the computation of $\cld(\cl_k, \cl_{g})$.
\cref{fig:update} illustrates one loop of the algorithm
where the entry being considered is $d' = \cld(\cl_i, \cl_\ell)\in
\cache_i$ (shown in the bottom gray box). 
In both cases, we store entry $d =
\cld(\cl_k, \cl_g) $ computed from $d'$ into both $\cache_k$ and
$\cache_g$ on Line~\ref{alg:updatecache:insert}.

Now we describe the update rule for computing $d$. For case (2), we can just directly apply the Lance-Williams formula~\cite{lance1967general} introduced in \cref{sec:prelims} 
and compute $\cld(\cl_i \cup \cl_j, \cl_g)$ from $\cld(\cl_i, \cl_g)$, $\cld(\cl_j, \cl_g)$, and $\cld(\cl_i, \cl_j)$.
For case (1), we can apply the Lance-Williams formula and compute $\cld(\cl_i \cup \cl_j, \cl_\ell \cup \cl_{\ell'})$
= $\cld(\cl_k, \cl_\ell \cup \cl_{\ell'})$ 
from $\cld(\cl_k, \cl_\ell)$, $\cld(\cl_k, \cl_{\ell'})$, and $\cld(\cl_\ell, \cl_{\ell'})$, where $\cl_{\ell'}$ is the cluster that $\cl_\ell$ merges with to form $\cl_g$. To compute $\cld(\cl_k, \cl_\ell)$ and $\cld(\cl_k, \cl_{\ell'})$, which are not cached since $\cl_k$ is a newly merged cluster, we can apply the Lance-Williams formula again since $\cl_k = \cl_i \cup \cl_j$. 
Below we give the update rule for average linkage with Euclidean distance metric as an example.
For case (2), where $\cl_g = \cl_\ell$, we have 
 \begin{align}\label{eq:case2}
  d = \frac{|\cl_i|}{|\cl_i|+|\cl_j|}  \cld_{\smb{avg-}1}(\cl_i, \cl_g) + \frac{|\cl_j|}{|\cl_i|+|\cl_j|} \cld_{\smb{avg-}1}(\cl_j, \cl_g).
\end{align}
For case (1), let $\cl_\ell$ and $\cl_{\ell'}$ be $\cl_g$'s children. We have 
 \begin{align}\label{eq:case1}
  d &= \frac{|\cl_i||\cl_\ell|}{|\cl_k||\cl_g|}  \cld_{\smb{avg-}1}(\cl_i, \cl_\ell) + \frac{|\cl_j||\cl_{\ell}|}{|\cl_k||\cl_g|} \cld_{\smb{avg-}1}(\cl_j, \cl_\ell) \nonumber \\
  &+\frac{|\cl_i||\cl_{\ell'|}}{|\cl_k||\cl_g|}  \cld_{\smb{avg-}1}(\cl_i, \cl_{\ell'}) + \frac{|\cl_j||\cl_{\ell'}|}{|\cl_k||\cl_g|} \cld_{\smb{avg-}1}(\cl_j, \cl_{\ell'}).
\end{align}
If the distances between $\cl_{\ell'}$ ($\cl_g$) and one or both of $\cl_i$ and $\cl_j$ 
are also cached in case (1) (case (2)), we can
also query them and accelerate the computation of $d$ by avoiding some
distance recomputation. For example, in update rule (\ref{eq:case2}), if $d' = \cld_{\smb{avg-}1}(\cl_i, \cl_g)$ is cached, we can compute $d=\cld_{\smb{avg-}1}(\cl_k, \cl_g)$ 
by $d = \frac{|\cl_i|}{|\cl_i|+|\cl_j|} d' + \frac{|\cl_j|}{|\cl_i|+|\cl_j|} \cld_{\smb{avg-}1}(\cl_j, \cl_g)$. If $\cld(\cl_j, \cl_g)$ is also cached, we can query for it and compute $d$ in constant time; otherwise, we need $|\cl_j||\cl_g|$ point distance computations to find $d$, which is less than the $|\cl_k||\cl_g|$
point distance computation required by a brute force method.

During the nearest neighbor range search, two clusters $\cl_i$ and $\cl_j$ might find each other as nearest neighbor candidates, and both want to compute $\cld(\cl_i, \cl_j)$ in parallel. 
A similar situation can happen when updating the cache entries for  $\cl_k$ and $\cl_g$.
Our implementation avoids these duplicate distance computations by having each cluster 
first insert a special entry into the hash table $\cache_{\min(i,j)}$ (or $\cache_{\max(i,j)}$ if $\cache_{\min(i,j)}$ is full), and then only compute the distance if the insertion was successful. The special entry can only be successfully inserted once for each pair of clusters
$\cl_i$ and $\cl_j$, and so $\cld(\cl_i, \cl_j)$ will only be computed once.

%% file: fig_exp_cachesize.tex
\begin{figure}[t]
\begin{center}
\includegraphics[width=\columnwidth]{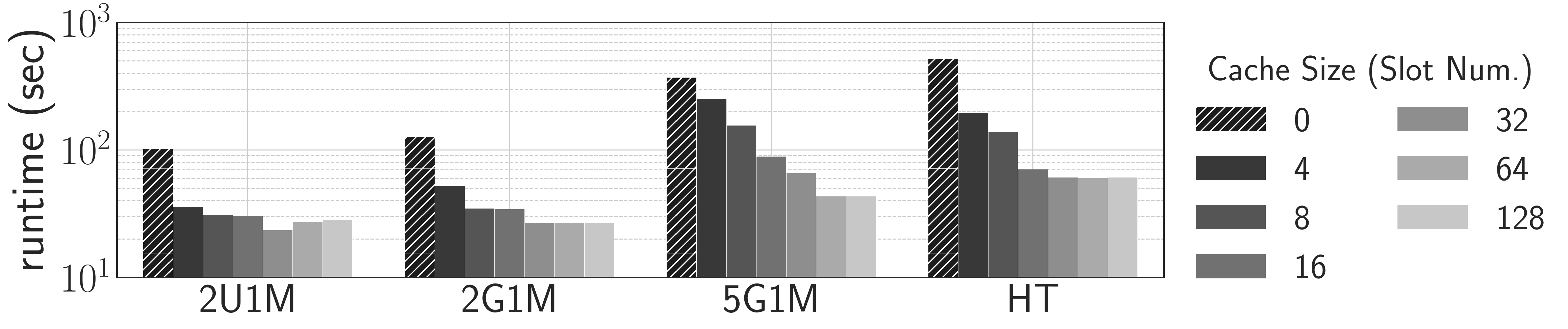}
\caption{Running times of using  \framework with average linkage and the Euclidean distance metric 
using 48-cores with two-way hyper-threading for varying cache sizes (values of $s$).
The data sets are labeled on the $x$-axis and are described in Section~\ref{sec:exp} (caption of \cref{table:compare}).
}
\label{plot:cachesize}
\end{center}
\end{figure}

%% file: pseudo_cache.tex
\begin{algorithm}[!t]
\DontPrintSemicolon
\fontsize{8pt}{8pt}\selectfont
\KwData{$\cl_k$ merged from $\cl_i$ and $\cl_j$, and distance structure $\distcomp$}
\tcp*[h]{$\cl$ is either $\cl_i$ (from $\cache_i$) or $\cl_j$ (from $\cache_j$)}\;
\PFor{$(d' = \cld(\cl, \cl_\ell)) \in \cache_i \cup \cache_j$ }{ \label{alg:updatecache:loop}
\lIf{$\set{\cl, \cl_\ell} == \set{\cl_i, \cl_j}$}{continue}\label{alg:updatecache:continue}
\eIf{$\cl_\ell$ is merged in this round}{  \label{alg:updatecache:1}
$\cl_g$ = the cluster that $\cl_\ell$ merged into\;
$d$ = $\distcomp$ computes $\cld(\cl_k, \cl_{g})$ from the distances among $\cl_i$, $\cl_j$, and the children of $\cl_g$ using $d'$ \; \label{alg:updatecache:1d}
}{ \label{alg:updatecache:2}
$\cl_g$ = $\cl_\ell$\;
$d$ = $\distcomp$ computes $\cld(C_k, C_{g})$ from the distances among $\cl_i$,  $\cl_j$, and $\cl_\ell$ using $d'$ \; \label{alg:updatecache:2d}
}
Insert $d$ into $\cache_k$ and $\cache_{g}$\;\label{alg:updatecache:insert}
}

\caption{Updating Cached Distance}\label{alg:updatecache}
\end{algorithm}

%% file: fig_cache.tex
\begin{figure}[t]
\centering
\includegraphics[trim=0 0 0 0, clip,width=0.99\columnwidth]{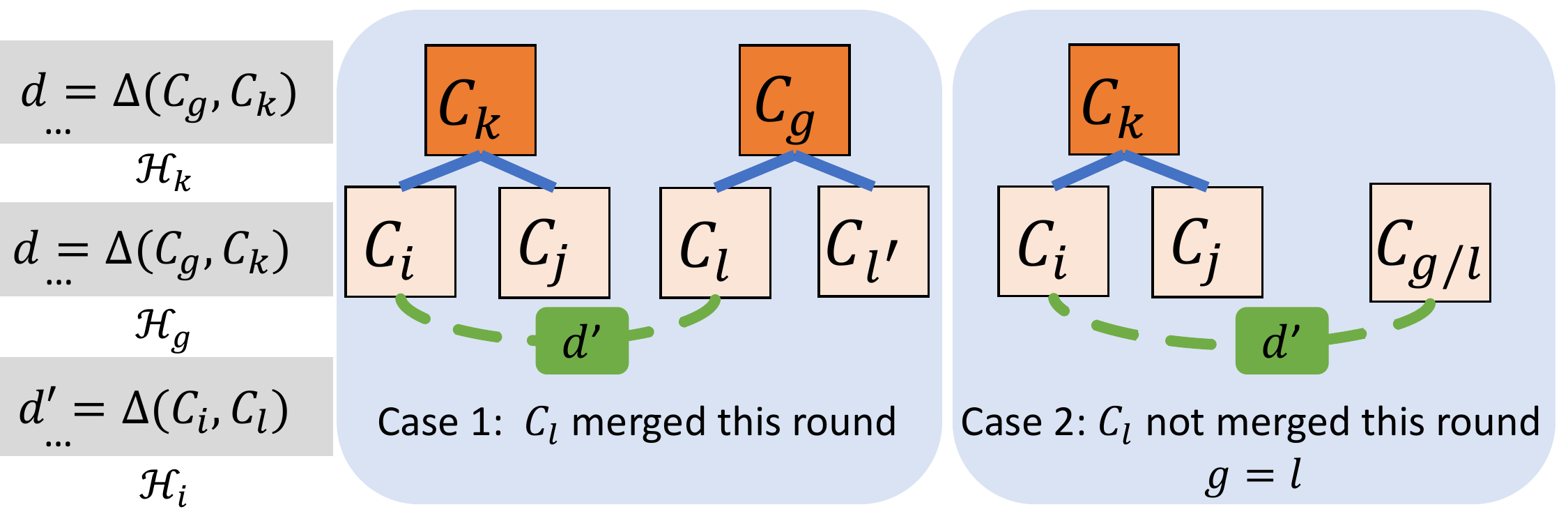}
\caption{An illustration of the cache table update in Algorithm~\ref{alg:updatecache}. The gray boxes show an entry in each of cache tables $\cache_k$, $\cache_g$, and $\cache_i$.
  The dark orange boxes are clusters merged in this round; the light orange boxes are clusters merged in previous rounds. The blue lines connect dendrogram children to their parent. The dotted green lines and boxes mark the cached distance between clusters. In case (1), $\cl_\ell$ is merged in this round into $\cl_g$; in case (2), $\cl_\ell$ is not merged in this round, and it is the same as $\cl_g$.   }
\label{fig:update}
\end{figure}

%% file: exp.tex
\section{Experiments}\label{sec:exp}

\myparagraph{Testing Environment} We perform experiments on
a \texttt{c5.24xlarge} machine on Amazon EC2, with $2$ Intel Xeon Platinum 8275CL (3.00GHz) CPUs for a total of 48
hyper-threaded cores, and 192 GB of RAM.  By default, we use all cores
with hyper-threading.  We use the \texttt{g++} compiler (version 7.5)
with the \texttt{-O3} flag, and use Cilk Plus, which is supported in
\texttt{g++}, for parallelism in our code~\cite{cilkplus}. For parallel experiments,
we use \texttt{numactl -i all} to balance the memory allocation across nodes. We also 
perform three runs of each parallel experiment and report the smallest running time. We
allocate a maximum of 15 hours for each run of a running time test, and do not report the
times for tests that exceed this limit.

We test the following implementations for HAC.
We refer to complete linkage as \defn{comp}, Ward's linkage as
\defn{Ward}, average linkage with Euclidean distance metric as
\defn{avg-1}, and average linkage with squared Euclidean distance
metric as \defn{avg-2}.

\begin{itemize}[topsep=1pt,itemsep=0pt,parsep=0pt,leftmargin=10pt]
\item \defn{PC} Our parallel \framework framework is implemented in C++, using the range query and caching optimizations.
  \item \defn{PC-mr} A parallel C++  
      NNC implementation that uses a distance matrix
      and merges all \rnns in each round. 
      All cluster distances are
      obtained from the distance matrix. 
      It uses the range query optimization to find the nearest neighbors.
	\item \defn{PC-m} A naive parallel C++ 
          NNC implementation that uses a distance matrix
          and merges all \rnns in each round. 
          All cluster distances are
          obtained from the distance matrix. 
          A parallel reduce is used over the distance matrix rows to find the nearest neighbor instead of
          using a range query.

	\item \defn{scipy (sc)}~\cite{mullner2011modern}  Scipy's serial implementation in Cython, which uses the NNC algorithm with a distance matrix for all of the linkage criteria tested. 
	\item \defn{sklearn (sk)}~\cite{scikit-learn} Scikit-learn's serial implementation in Cython, which uses a distance matrix. It has a heap for all distances and merges the global closest neighbor pair on each round. 
	\item \defn{fastcluster}~\cite{mullner2013fastcluster} A serial C++ implementation of HAC with a Python interface. It contains two approaches of implementations of HAC---one
	generic implementation (\defn{fc-gen}) uses the naive algorithm where the global \rnn pair is merged in each round, and the other (\defn{fc-nnc})
	is based on the NNC algorithm using a distance matrix.

  For Ward's linkage, fastcluster has a linear space implementation for the naive method that work by computing the cluster distances on the fly using the cluster centroids.
        We also wrote linear-space implementations for the NNC algorithm that compute distances on the fly for Ward's linkage and average linkage with squared Euclidean distance metric.
  We report the running time of the linear space implementation when available. 
	\item \defn{fastprotein (fp)}~\cite{hung2014fast_protein_cluster} A parallel C++ implementation that parallelizes the naive NNC algorithm by computing the global \rnn and updating the distance matrix in parallel on each of the $n-1$ rounds. 
	It only supports complete and average linkage.
	\item \defn{Jeon (Je)}~\cite{jeon2014multi} A parallel C++ implementation of the parallel \nnc algorithm by Jeon and Yoon.~\cite{jeon2014multi}, which only supports average linkage with the Euclidean distance metric.
	\item \defn{Althaus (Al)}~\cite{althaus2014greedy} Our parallel C++ implementation of Althaus et al.'s complete
  linkage algorithm that uses linear memory.
\end{itemize}

\myparagraph{Data Sets} We use both synthetic and real-world datasets, all of which fit in the RAM of our machine. Let $n$ be the number of points.
The \defn{GaussianDisc} data set
contains points inside a bounding hypergrid with side length
$5\sqrt{n}$.  $90\%$ of the
points are equally divided among five clusters, each with a Gaussian
distribution.  Each cluster has its mean randomly sampled from the hypergrid, a standard deviation of $1/6$, and a diameter of
$\sqrt{n}$. The remaining points are randomly distributed.
The \defn{UniformFill} data set contains points distributed uniformly
at random inside a bounding hypergrid with side length $\sqrt{n}$.  We generate the synthetic
data sets with 10 million points for dimensions $d=2$
and $d=5$. 

We also use two existing simulation datasets.
\defn{UCI1}~\cite{bock2004methods} is a 10-dimensional data
set with $19,020$ data points. This data set is generated to simulate
registration of high energy gamma particles~\cite{heck1998monte}.
\defn{UCI4}~\cite{keogh1999indexing} is a 10-dimensional data
set with $100,000$ data points. This data set has a pseudo-periodic
time series for each of its dimension, and hence is likely to form
long chains.

We use the following real-world data sets.
\defn{GeoLife}~\cite{Zheng2008} is a 3-dimensional data
set with $24,876,978$ data points. This data set contains user
location data, and is extremely skewed.
\defn{HT}~\cite{Huerta2016OnlineHA} is a 10-dimensional data
set with $928,991$ data points containing home sensor data.
\defn{CHEM}~\cite{fonollosa2015reservoir} is a 16-dimensional
data set with $4,208,261$ data points containing chemical sensor data.

When referring to the data sets in this section,
we use a prefix to indicate its dimensionality and suffix to indicate its size.
To obtain smaller data sets, we randomly sample from the corresponding
larger data sets. The letter "U" indicates UniformFill and "G"indicates "GaussianDisc".

\myparagraph{Cache Sizes}
In our experiments, we use a cache size of $s=64$ for
avg-1 and $s=0$ for complete,
Ward, and avg-2 except otherwise noted. The choice of $s=64$
will be discussed in more detail in Section~\ref{sec:expopt} 
along with the benefit of our range query
and caching optimizations. We use $s=0$ for complete,
Ward, and avg-2 to show our framework achieves good 
performance on cheaper linkage criteria even without caching.

\subsection{Comparison with Other Implementations}

\input{fig_exp1}

\cref{plot:compare} shows the running times vs.\ number of threads
for all of the serial and parallel implementations on three small data
sets (2D-GaussianDisc-10K, 10D-UCI1-19K, and 10D-UCI4-100K).
Implementations with a single data point are
serial. We only compare them on the small data
sets because the algorithms that require quadratic memory run out of
memory for larger data sets.

We see that our implementation PC almost always
outperforms existing implementations across all thread counts.  Even
the version of our algorithm using the distance matrix without the range query optimization
(PC-m) is faster than all other implementations at higher thread
counts.  Unlike the existing parallel implementations, fastprotein (fp) and
Jeon (Je), our implementations are more scalable since we merge all \rnn
pairs on each round, and do not use locks. On the small data sets, using 48 cores with hyper-threading, PC
is 5.8--88.0x faster than fp and 37.5--110.1x faster than Je.
Table~\ref{table:compare} shows the running times for PC,
fc-gen, and fc-nnc on larger data sets (Je does not scale to these data
sets due to its quadratic memory requirement), and we see that PC is
64.77--733.90x faster than fastcluster on these data sets.  On a single thread,
we find that PC is 2.19--47.92x  faster than the next fastest implementation
(except on 10D-UCI1-19K for complete linkage, where PC is 1.58x slower than fc-nnc and fc-gen).

In Figure~\ref{plot:compare}, PC shows limited scalability on higher
thread counts, because these data sets are small and the overhead of
using more threads is high relative to the work of the algorithm.
However, in the next subsection, we show that PC is able to achieve
higher parallel scalability on larger data sets.

\subsection{Scalability}\label{sec:scalability}

\myparagraph{Scalability with Thread Count} Table~\ref{table:compare}
and Figure~\ref{plot:exp_scal} present the runtime and scalability of
PC on different numbers of threads for larger data sets, which
most existing implementations do not scale to.  For average
linkage with the Euclidean distance metric, the speedups for several
data sets are not shown since the single-threaded experiments timed
out.  We see that using 48 cores with two-way hyper-threading, PC achieves
15.42--54.23x speedups on complete linkage, 15.43--44.16x speedups on
Ward's linkage, 31.95--45.54x speedups on average linkage with
Euclidean distance, and 13.75--46.6x speedups on average linkage using
squared Euclidean distance.
From Figure~\ref{plot:exp_scal}, we can see that on most data sets,
our algorithm keeps scaling up until 48 threads.

\input{fig_scalability_data}
\myparagraph{Scalability with Data Size}
Figure~\ref{plot:exp_scal_data} shows the runtimes of our algorithm
PC on three data sets of varying sizes using 48 cores
with two-way hyper-threading. We observe that  PC scales well
with data set size. The scalability is better for comp, Ward, and avg-2
than for avg-1 because avg-1 always requires quadratic work to compute cluster distances, while Ward and avg-2 require constant time to compute them and comp usually requires less than quadratic time to compute them due to pruning.

\input{fig_exp_decomp}
\myparagraph{Runtime Decomposition} 
We now
describe the breakdown of running time across different steps of
\framework, as well as the scalability of each step.
Figure~\ref{plot:decomp} shows the speedups and running times of steps of ~\mbox{avg-1} using 48 cores with hyper-threading on different data sets.  
From \cref{alg:framework}, ``init'' corresponds to
Lines~\ref{alg:framework:initstart}--\ref{alg:framework:buildtree}
where we initialize the data structures; ``nn'' corresponds to
Lines~\ref{alg:framework:fnn}--\ref{alg:framework:chain2} where we
find the nearest neighbors of all terminal nodes and update the
chains; ``merge'' corresponds to
Lines~\ref{alg:framework:findrnn}--\ref{alg:framework:cache} where we
merge the \rnns and update the cache tables if $s>0$; and ``update''
corresponds to
Lines~\ref{alg:framework:dist}--\ref{alg:framework:updateterm} where
we update the data structures to prepare for next round.

From Figure~\ref{plot:decomp} (top), we see that the ``nn'' and ``merge'' steps are more scalable with respect to thread count than 
the ``update'' and ``init'' steps. The reason is that in the ``nn'' step, we
find the nearest neighbor of all terminal nodes in parallel, and in the ``merge'' step, we 
merge all \rnns in parallel. The numbers of
terminal nodes and \rnns are usually much larger than the number of available
threads, and thus there is a lot of opportunity for parallelism. The ``init'' and ``update'' step are less scalable because they have less work to be divided across threads.

From Figure~\ref{plot:decomp} (bottom), we see that the ``nn'' and ``merge'' steps are less scalable with respect to data size than 
the ``update'' and ``init'' steps. This is because the ``nn'' and ``merge'' steps asymptotically dominate the work of the whole algorithm.

\subsection{Analysis of Our Framework}\label{sec:expopt}
We now discuss the effects of our range query and caching optimizations and show our running time is close to quadratic in practice.
From \cref{plot:compare},
PC-mr is 1.67x faster on average than PC-m on 48 cores with two-way hyper-threading, which
shows that the benefit of using our optimized range query is larger than its overhead
even on these small datasets. PC is 15.06x faster on average than PC-mr because
although PC needs to compute some distances on the fly, PC avoids the overhead of 
computing the distance matrix and updating the matrix in each round.
This shows the benefit of avoiding a distance matrix.

To further show the benefit of our range query and caching optimizations, we measure the maximum average cache usage 
$\max_{r \in \text{rounds}} \left( \underset{C \in \text{clusters}}{\text{avg}} (\text{\# of cache slots used by } C \text{ in round }r)\right)$
for avg-1 using $s = 256$  for all clusters (we used $s=128$ for GeoLife due memory limitations). We use a larger cache 
size than our previous experiments so that fewer clusters hit the size limit, which gives us a more accurate analysis of cache usage. 
We found that the maximum average cache usage ranges from 6.7--84.5 slots.
This explains why runtimes stop decreasing for cache sizes larger than 64 in \cref{plot:cachesize}. If the cache is too large and many of the entries are empty, the runtime could be slower than using a smaller cache because when we merge the caches, we need to filter out the non-empty entries, and thus larger caches incur more overhead. 

We use $s=64$ for avg-1 in all other experiments to show that our framework can achieve good performance with a relatively small cache so that our memory overhead is minimal. 
\cref{plot:cachesize} shows that $s=64$ gives good performance across different data sets.
\cref{sec:memory} shows our memory usage is very small
with $s=64$.

\input{fig_exp_stats}

We now show that our running time is close to quadratic in practice. 
As described in \cref{sec:framework}, the work of our framework is bounded by $M = \sum |\clusters_i| (|\terminal_i|+\log |\clusters_i|)$
plus the cost of distance computations, $D$.
\cref{plot:exp_opt} (left) shows that $M$ and $D$ are quadratic in the number of points in practice. \cref{plot:exp_opt} (right) shows that only a very small fraction of clusters are included in the range queries and an even smaller fraction of distance computations between clusters are required in practice.

\input{fig_scalability_memory}
\subsection{Memory Usage}\label{sec:memory}
\cref{table:memory} shows the memory usage (in megabytes) vs.\ data set size for 2D-GaussianDisc data sets for the different implementations. We measure the memory usage using the Valgrind Massif heap profiler tool~\cite{nethercote2007valgrind}.
For PC and PC-m, we use 48 cores with two-way hyper-threading. For fastprotein (fp) we use 36 threads, for Jeon (Je) we use 4 threads, and for Althaus (Al) we use 36 threads. The number of threads are chosen for best performance based on Figure~\ref{plot:compare}.

Our algorithm PC shows linear memory increase vs.\ data size  on all methods while most other methods, except Al, fc-gen for Ward, and fc-nnc
for Ward and avg-2 (which require linear space), show quadratic memory increase vs.\ data size, which is consistent with the fact that they use a distance matrix.
PC uses less memory than all other implementations, except that Al uses less memory for complete linkage, 
and Je uses less memory for the small 1K data set for avg-1.
However, Al only works for complete linkage and is orders of magnitude slower than PC (\cref{plot:compare}). PC uses less than 2x of the memory used by Al. Overall, PC uses up to 237.3x less memory than existing implementations.

%% file: fig_exp1.tex
\begin{figure*}[t]
\begin{center}
\includegraphics[width=\textwidth]{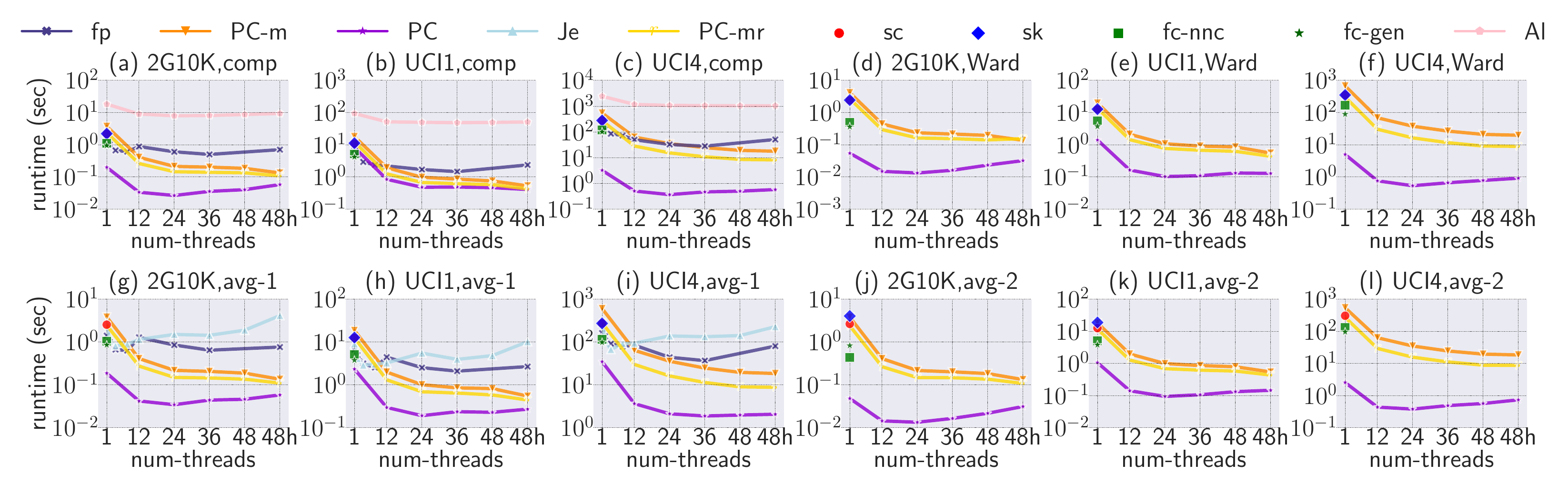}
\caption{Runtimes (seconds) of our algorithms compared with other implementations with varying thread counts. (48h) indicates 48 cores with two-way hyper-threading. Implementations with a single data point are serial. Je only supports average linkage with Euclidean distance and fp only supports complete and average linkage with Euclidean distance. sc and sk have very similar running times and overlap on some plots. For avg-2, sk runs out of memory for the UCI4 data set. Our algorithm PC is faster than all other implementations for all number of threads, except on a single thread for the UCI1 data set using complete linkage. See Table~\ref{table:compare} and Figure~\ref{plot:exp_scal} for running times and scalability of PC on larger data sets. 
}
\label{plot:compare}
\end{center}
\end{figure*}

\begin{table*}[t]
  \caption{Runtimes (seconds) and self-relative parallel speedups of fastcluster and PC. ``PC-1'' is our runtime on 1 thread and ``PC-48h'' is our runtime using 48 cores with two-way hyper-threading. PC on avg-1 timed out on several larger data sets.
    For fastcluster on several larger data sets, we only have runtimes for Ward and avg-2 because it requires quadratic memory for comp and avg-1 and runs out of memory. For Ward, we report fc-gen because it is faster than fc-nnc. "--" means that the running time exceeds 15 hours. For the data set names, the first number indicates dimension, the letter "U" indicates UniformFill, "G" indicates "GaussianDisc", "1M" indicates 1 million data points, and "10M" indicates 10 million data points.
  }\label{table:compare}
  \begin{center}
  \begin{small}
  \begin{tabular}{c@{}c c@{  }@{  }@{  }c@{  }@{  }@{  }c@{  }@{  }@{  }c@{  }@{  }@{  }c@{  }@{  }@{  }@{  }c@{  }@{  }@{  }c@{  }@{  }@{  }c@{  }@{  }@{  }c@{  }@{  }c@{  }@{  }@{  }c}
  \toprule
  & & 2U1M & 2G1M & 5U1M & 5G1M & HT & 2U10M & 2G10M & 5U10M & 5G10M & CHEM & Geolife \\[0.5 ex]
      \hline
  \multirow{3}{*}{ comp \phantom{abc} } & PC-1 (sec)  \phantom{abc}& 47.21 & 50.47 & 2119.10 & 2282.00 & 56.80 & 677.88 & 609.08 & 53008.00 & 49052.00 & 4972.90 & 1948.90 \\
  & PC-48h (sec) \phantom{abc} & 1.34 & 1.61 & 40.40 & 46.74 & 3.68 & 19.01 & 21.29 & 977.47 & 1038.82 & 126.54 & 123.34 \\
  & self-speedup  \phantom{abc}& 35.12 & 31.39 & 52.45 & 48.83 & 15.42 & 35.66 & 28.61 & 54.23 & 47.22 & 39.30 & 15.80 \\\hline
  \multirow{4}{*}{ Ward  \phantom{abc}} & fc-gen (sec) \phantom{abc} & 3537.11 & 3845.04 & 7284.11 & 6676.75 & 8760.04 & -- & -- & -- & -- & -- & -- \\
  & PC-1 (sec)  \phantom{abc}& 11.31 & 12.23 & 79.11 & 103.09 & 23.13 & 175.70 & 156.53 & 1475.50 & 1230.30 & 1096.40 & 681.63 \\
  & PC-48h (sec) \phantom{abc} & 0.50 & 0.52 & 1.95 & 2.69 & 1.50 & 5.69 & 5.20 & 30.36 & 34.99 & 24.83 & 32.14 \\
  & self-speedup \phantom{abc} & 22.42 & 23.70 & 40.52 & 38.36 & 15.43 & 30.90 & 30.11 & 48.59 & 35.17 & 44.16 & 21.21 \\\hline
  \multirow{3}{*}{ avg-1  \phantom{abc}} & PC-1 (sec)  \phantom{abc}& 859.22 & 857.07 & 1627.80 & 1734.10 & 2652.70 & -- & -- & -- & -- & -- & -- \\
  & PC-48h (sec)  \phantom{abc}& 23.59 & 26.83 & 38.10 & 45.27 & 58.25 & 2969.93 & 3206.48 & 6323.56 & 5772.03 & 2323.38 & 19213.60 \\
  & self-speedup  \phantom{abc}& 36.42 & 31.95 & 42.72 & 38.30 & 45.54 & -- & -- & -- & -- & -- & -- \\\hline
  \multirow{4}{*}{ avg-2  \phantom{abc}} & fc-nnc (sec)  \phantom{abc}& 4602.83 & 4022.04 & 8907.50 & 11425.99 & 13244.60 & -- & -- & -- & -- & -- & -- \\
  & PC-1 (sec)  \phantom{abc}& 10.28 & 11.11 & 62.10 & 81.21 & 18.05 & 159.65 & 141.52 & 1146.20 & 955.77 & 833.06 & 575.39 \\
  & PC-48h (sec) \phantom{abc} & 0.47 & 0.49 & 1.65 & 2.27 & 1.31 & 5.30 & 4.80 & 24.60 & 28.51 & 20.93 & 28.34 \\
  & self-speedup  \phantom{abc}& 21.78 & 22.76 & 37.64 & 35.73 & 13.75 & 30.14 & 29.49 & 46.60 & 33.52 & 39.79 & 20.30 \\
  \bottomrule
  \end{tabular}
  \end{small}
  \end{center}
  \end{table*}

\begin{figure*}[t]
\begin{center}
\includegraphics[width=\textwidth]{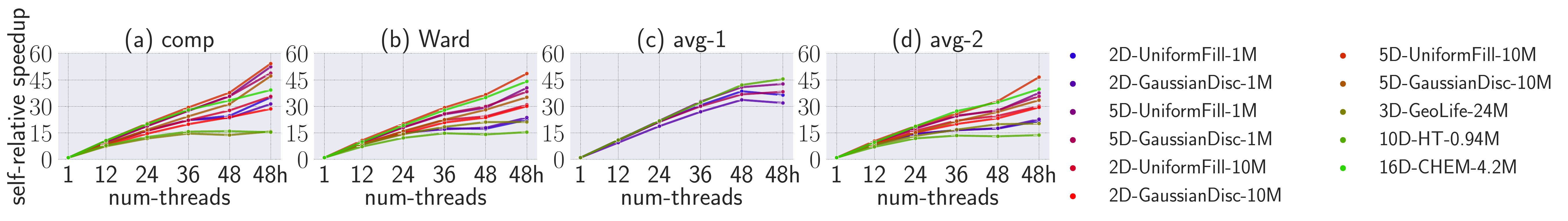}
\caption{Self-relative parallel speedup vs.\ thread counts for complete, Ward's, and average linkage using our ``PC'' algorithm on large datasets.
(48h) indicates 48 cores with two-way hyper-threading.
  For average linkage with the Euclidean distance metric, the speedups for several data sets are not shown since the single-threaded experiments timed out.
}
\label{plot:exp_scal}
\end{center}
\end{figure*}

%% file: fig_scalability_data.tex
\ifarxiv
\begin{figure*}[t]
\begin{center}
\includegraphics[width=\textwidth]{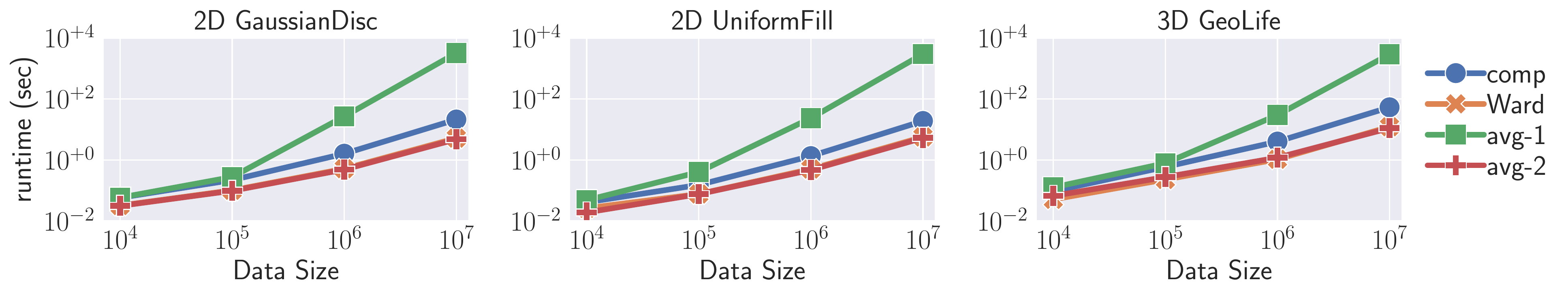}
\caption{Runtimes (seconds) of our algorithm ``PC'' on three data sets of varying sizes using 48 cores with two-way hyper-threading. The ``Ward'' and ``avg-2'' lines overlapped because they have similar runtimes. The results show that the scalability is very similar across datasets.
}
\label{plot:exp_scal_data}
\end{center}
\end{figure*}

\else

\begin{figure}[t]
\begin{center}
\includegraphics[width=\columnwidth]{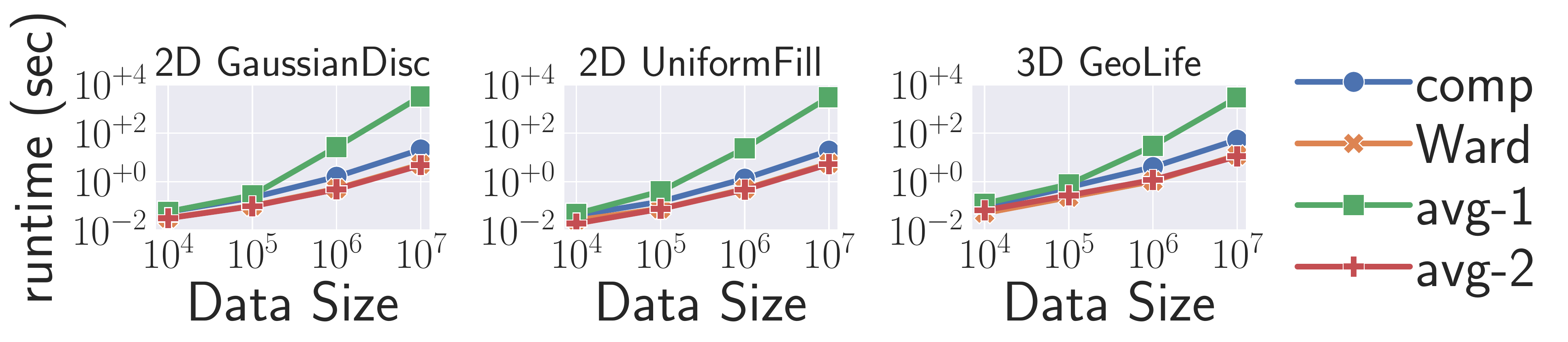}
\caption{Runtimes (seconds) of our algorithm ``PC'' on three data sets of varying sizes using 48 cores with two-way hyper-threading. The ``Ward'' and ``avg-2'' lines overlapped because they have similar runtimes. The results show that the scalability is very similar across datasets.
}
\label{plot:exp_scal_data}
\end{center}
\end{figure}
\fi

%% file: fig_exp_decomp.tex
\ifarxiv
\begin{figure*}[t]
  \begin{center}
  \includegraphics[width=\textwidth]{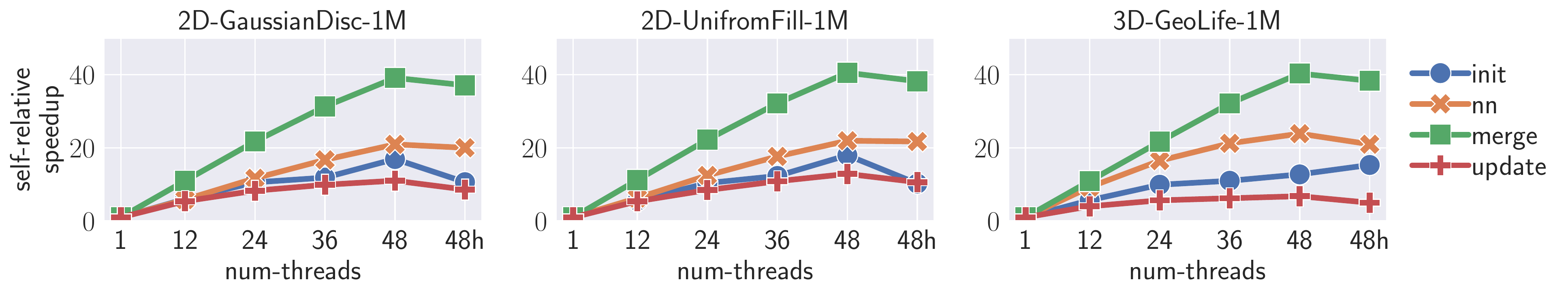}
  \includegraphics[width=\textwidth]{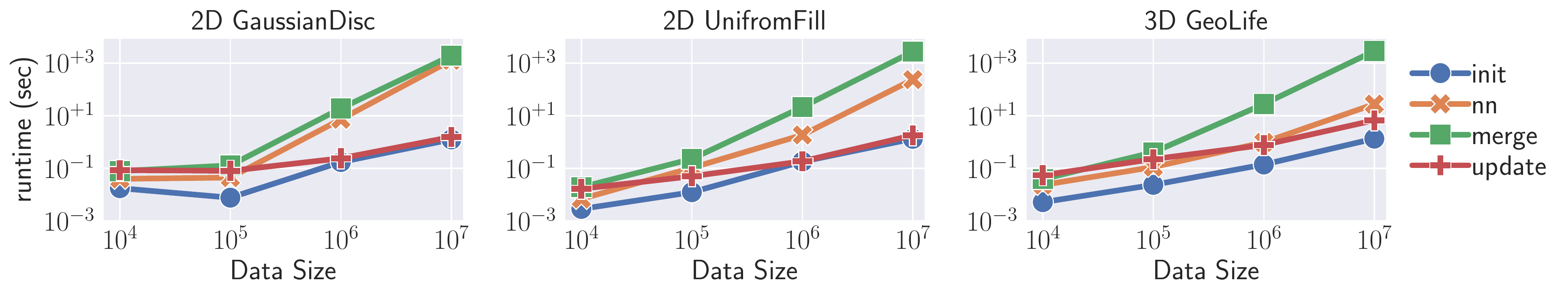}
  \caption{Self-relative parallel speedup and runtimes (seconds) of steps of avg-1 using 48 cores with
  two-way hyper-threading on different thread counts and data set sizes. 
  From
    \cref{alg:framework}, "init" corresponds to
    Lines~\ref{alg:framework:initstart}--\ref{alg:framework:buildtree};
    "nn" corresponds to
    Lines~\ref{alg:framework:fnn}--\ref{alg:framework:chain2}; "merge"
    corresponds to
    Lines~\ref{alg:framework:findrnn}--\ref{alg:framework:cache}; and
    "update" corresponds to
    Lines~\ref{alg:framework:dist}--\ref{alg:framework:updateterm}. 
      }
  \label{plot:decomp}
  \end{center}
  \end{figure*}
  
\else

\begin{figure}[t]
  \begin{center}
  \includegraphics[width=\columnwidth]{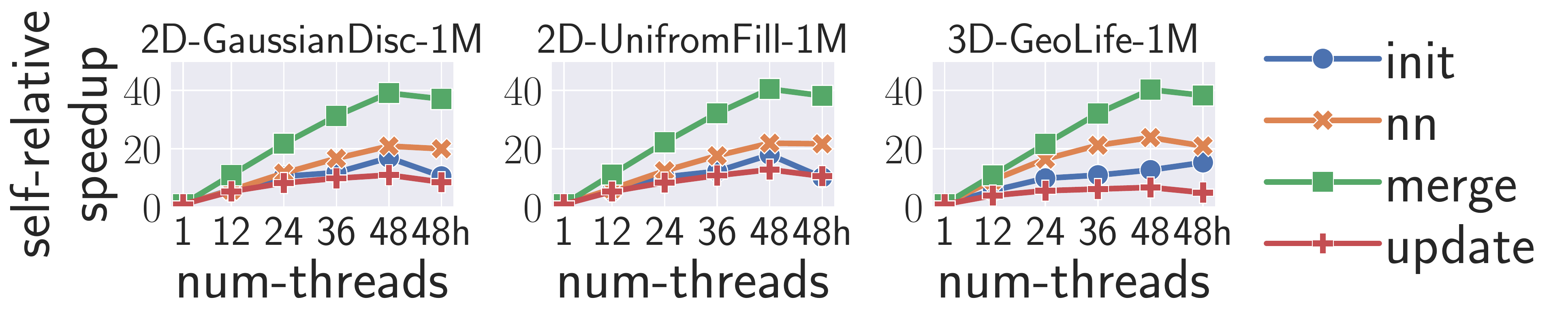}
  \includegraphics[width=\columnwidth]{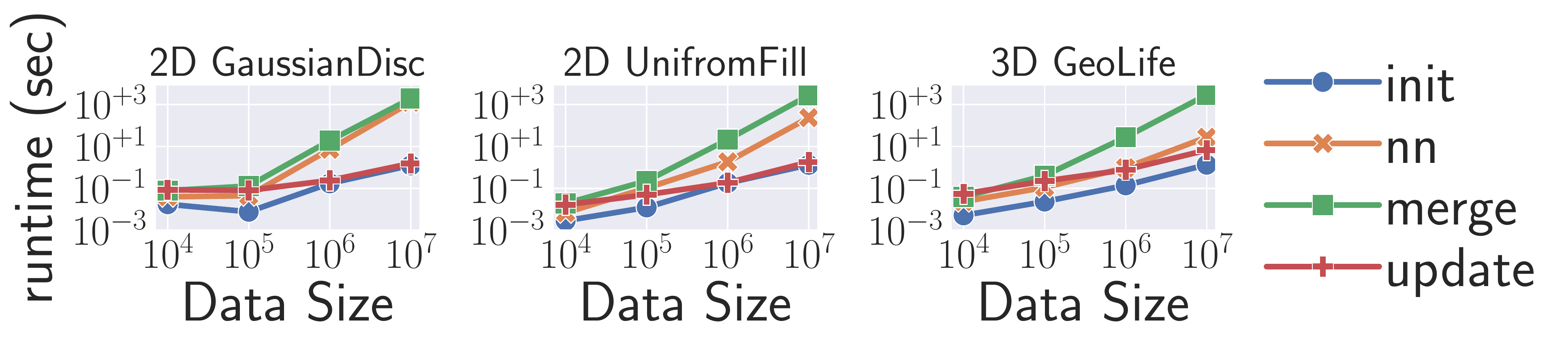}
  \caption{Self-relative parallel speedup and runtimes (seconds) of steps of avg-1 using 48 cores with
  two-way hyper-threading on different thread counts and data set sizes. 
  From
    \cref{alg:framework}, "init" corresponds to
    Lines~\ref{alg:framework:initstart}--\ref{alg:framework:buildtree};
    "nn" corresponds to
    Lines~\ref{alg:framework:fnn}--\ref{alg:framework:chain2}; "merge"
    corresponds to
    Lines~\ref{alg:framework:findrnn}--\ref{alg:framework:cache}; and
    "update" corresponds to
    Lines~\ref{alg:framework:dist}--\ref{alg:framework:updateterm}. 
      }
  \label{plot:decomp}
  \end{center}
  \end{figure}
\fi

%% file: fig_exp_stats.tex
\begin{figure}[t]
  \begin{center}
  \includegraphics[width=\columnwidth]{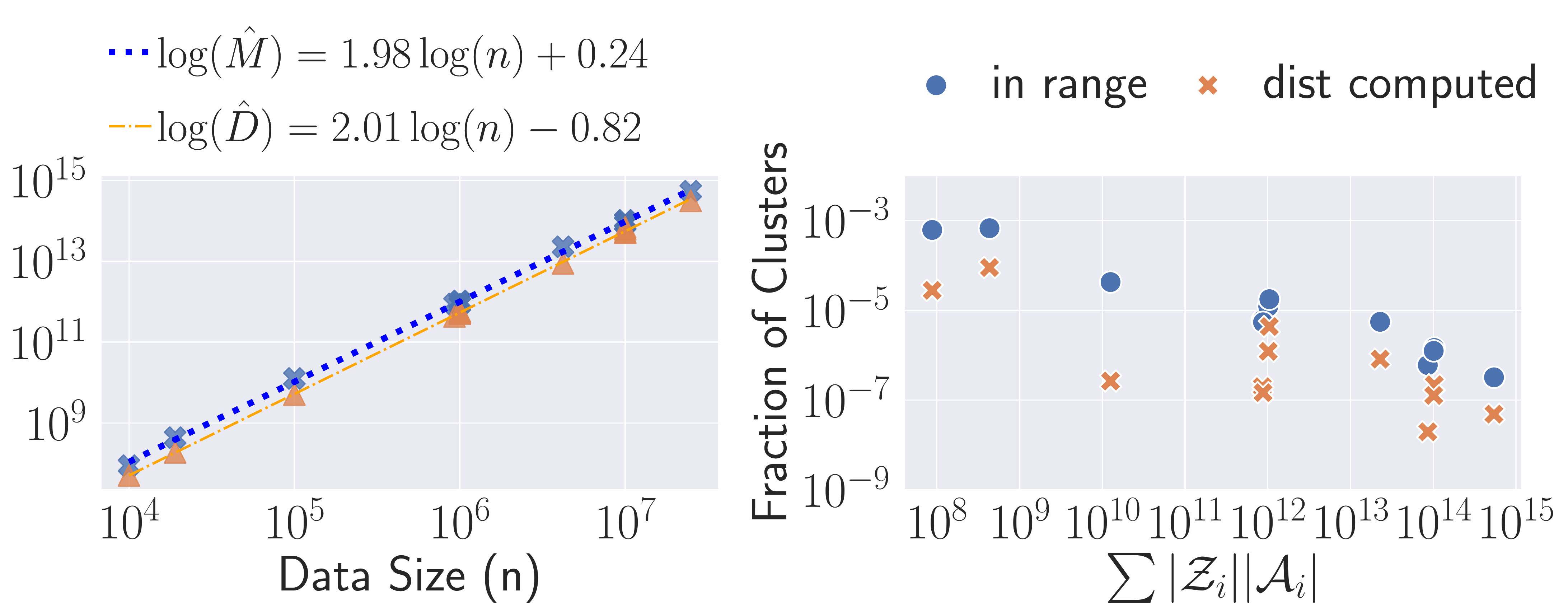}
  \caption{Statistics for avg-1 on different datasets with $s=64$. (Left)
  Blue crosses denotes $M = \sum (|\terminal_i| +\log |\clusters_i|)|\clusters_i|$ for different data sets. 
  Orange triangles denote the total number of distance computations between points ($D$). $\hat{M}$ and $\hat{D}$ are least squares fitted lines of $M$ and $D$, respectively.
  (Right) Fraction of clusters in the range queries and fraction of distance computations required between clusters. 
  }
  \label{plot:exp_opt}
  \end{center}
  \end{figure}

%% file: fig_scalability_memory.tex
\begin{table}[t]
  \begin{center}
  \caption{Memory usage (MB) vs.\ data set size for 2D-GaussianDisc data sets for the different implementations. The smallest memory for each linkage criteria and data set is in bold.
}\label{table:memory}
  \begin{small}
    \begin{tabular}{cccccccccccc}
      \toprule
      & n & fc-gen & fc-nnc & PC & PC-m & PC-mr & sc & sk & fp & Je & Al \\ \hline
      \multirow{3}{*}{comp} & 1K &17.5& 17.5 & 5.8 & 8.5 & 8.7 & 21.6 & 31.4 & 4.2 & --&\textbf{3.40}\\
      & 3K &49.5& 49.5 & 11.1 & 41.4 & 42.1 & 85.8 & 95.5 & 36.4 & --& \textbf{7.29}\\
      & 10K &414.6& 414.4 & 27.6 & 407.7 & 410.2 & 814.2 & 824.0 & 401.1 & --& \textbf{20.75} \\\hline
      \multirow{3}{*}{Ward} & 1K &15.4& 15.4 &\textbf{ 3.5} & 8.6 & 8.7 & 21.6 & 31.4 & -- & -- & --\\
      & 3K &20.2& 20.2 & \textbf{4.8} & 41.4 & 41.9 & 85.8 & 95.5 & -- & --& -- \\
      & 10K &27.6& 27.6 & \textbf{9.2} & 407.7 & 409.3 & 814.2 & 824.0 & -- & --& -- \\\hline
      \multirow{3}{*}{avg-1} & 1K&17.5 & 17.5 & 6.0 & 8.5 & 8.7 & 21.6 & 31.4 & 3.2 & \textbf{3.0}& -- \\
      & 3K &49.5& 49.5 & \textbf{12.5} & 41.3 & 42.1 & 85.8 & 95.5 & 21.3 & 20.7 & --\\
      & 10K &414.7& 414.5 & \textbf{32.7} & 407.7 & 410.2 & 814.2 & 824.0 & 210.7 & 208.7 & --\\\hline
      \multirow{3}{*}{avg-2} &1K& 17.5 & 15.4 & \textbf{3.6} & 8.6 & 8.7 & 21.6 & 47.4 & -- & -- & --\\
      & 3K &49.5& 20.2 & \textbf{5.1} & 41.3 & 42.1 & 85.8 & 239.5 & -- & -- & --\\
      & 10K &414.7& 27.6 & \textbf{10.2} & 407.7 & 410.2 & 814.2 & 2423.9 & -- & -- & --\\
      \bottomrule
      \end{tabular}
      
\end{small}
\end{center}
\vspace{-3 pt}
\end{table}

%% file: related_work.tex
\section{Related Work}\label{sec:related}

There is a rich literature in designing HAC algorithms.  In the most
naive algorithm, a distance matrix is used to maintain all pairwise
distances between clusters. On each iteration, the matrix is searched
to find the closest pair of clusters, which are then merged, and
distances to this newly merged cluster are computed.  The algorithm
runs for $n-1$ iterations, after which a single cluster remains.  A
straightforward implementation of this algorithm gives $O(n^3)$
time, but it can be improved to $O(n^2 \log n)$ time by storing matrix
entries in heap-based priority queues~\cite{olson1995parallel,
  mullner2013fastcluster}.

The two popular Python libraries
scipy~\cite{Virtanen2020scipy} and
scikit-learn~\cite{scikit-learn} both provide sequential
algorithms for HAC.  The two libraries' implementations both compute
and store a distance matrix.  
Fastcluster~\cite{mullner2013fastcluster} contains
three implementations of HAC---two heap-based naive algorithms for
general linkage functions, where one uses the distance matrix and the
other compute cluster distances on the fly, and an NNC algorithm that
uses the distance matrix.  Lopez-Sastre et al.~\cite{fastrnn2012}
propose a sequential NNC algorithm that speeds up the chain
construction using a dynamic slicing strategy that only searches for
the nearest neighbor within some slices. Their algorithm only works
for linkage functions where the distance can be expressed using
centroids and variances.

 There have been  implementations that focus on reducing
the in-memory space usage of HAC from quadratic to linear by writing the
quadratic-space distance matrix to disk, and loading it into memory in smaller
chunks~\cite{nguyen2014sparsehc,loewenstein2008efficient}.
These algorithms are
sequential, and only merge one pair of clusters at a time. In contrast, our algorithm is parallel, and also
does not require writing or loading additional information to and from
disk.
Moreover, the algorithms above are designed to take advantage of sparse distance metrics,
where only some distances between data points are defined while other
distances are considered to be "missing" and the points have
"large" dissimilarity between them, making them less suitable for the Euclidean distance or squared Euclidean distance metrics.

There have also been many parallel algorithms developed for HAC,
although it is difficult to parallelize in
theory~\cite{greenlaw2008parallel}.  Olson~\cite{olson1995parallel}
gives parallel algorithms, some of which parallelize the
NNC algorithm by finding the nearest neighbor in parallel on each
round, but  still only merges one pair per round, and so
there will always be $n-1$ rounds.  Li~\cite{li1990parallel} gives
parallel HAC algorithms that store the distance matrix, based on an
older theoretical model for a SIMD machine with distributed memory.
Li and Fang~\cite{li1989parallel} give parallel HAC algorithms on
hypercube and butterfly network topologies.  Du and
Lin~\cite{du2005novel} give a parallel HAC algorithm on a cluster of
compute nodes.  Zhang et al.~\cite{zhang2019dhc} propose a distributed
algorithm for HAC that partitions the datasets using \kdt{s} or
quadtrees, and then for each leaf node, finds a region where the \rnn pairs
might exist. In parallel, each compute node finds the local \rnn pairs in a
region, and then global \rnn pairs are found from the local pairs.
This
method merges multiple \rnn pairs, but
their paper does not specify how the distances between clusters are
updated or computed after merges.
Fastprotein~\cite{hung2014fast_protein_cluster} is a naive
parallelization of fastcluster.  
Sun et al.~\cite{sun2009esprit} develop a parallel version of algorithms that write the distance matrix to disk and load chunks of it into memory~\cite{nguyen2014sparsehc,loewenstein2008efficient}. However, they still only merge one pair of clusters at a time.
Jeon and Yoon~\cite{jeon2014multi} present a parallel NNC algorithm
using a distance matrix, which we discussed earlier.
Althaus et al.~\cite{althaus2014greedy} present a parallel complete
linkage algorithm that uses linear main memory; however their
algorithm requires $n-1$ rounds because they only merge the global
\rnn pair on each round.  
Sumengen et al.~\cite{sumengen2021scaling} developed 
a distributed algorithm for HAC that 
recovers the exact HAC solution, and has good theoretical bounds
under assumptions about the underlying cluster structure of the data.
Similar to our algorithm, it merges reciprocal nearest-neighbor pairs in parallel.
Their algorithm can be accelerated in practice by using a graph-building 
approach, and has good empirical performance on billion-scale datasets.
In contrast, our work is focused on designing an optimized shared-memory framework for
running HAC on the original points and the $O(n^2)$ distances between them, 
while accelerating the distance computation steps and distance storage steps. 
It would be interesting to compare both algorithms in a shared-memory setting 
in future work.

Besides the linkage criteria considered in this paper, other popular
criteria for HAC include single, centroid, and median
linkage. Single linkage with the Euclidean metric is closely related to
the Euclidean minimum spanning tree problem, and can be solved
efficiently using variants of minimum spanning
tree algorithms~\cite{wang2021fast, march2010fast}. Centroid and median linkage
do not satisfy the reducibility property and cannot take advantage of
the NNC algorithm.
There has also been work on
 other hierarchical clustering methods, such as partitioning
hierarchical clustering algorithms and algorithms that combine
agglomerative and partitioning
methods~\cite{li1990parallel,rajasekaran2005efficient,dash2004efficient,monath2020scalable,cathey2007exploiting}.
Finally, there has been work on analyzing the cost function of the HAC problem~\cite{moseley2017approximation,
dasgupta2005performance, dasgupta2016cost, cohen2019hierarchical, dhulipala2021hierarchical} and approximating the HAC problem on
various linkage criteria and metrics~\cite{cochez2015twister,
  monath2019scalable, kull2008fast, gilpin2013efficient,
  bateni2017affinity, defays1977efficient, krznaric2002optimal, 
  abboud2019subquadratic, chatziafratis2020hierarchical}.

%% file: conclusion.tex
\section{Conclusion}\label{sec:conclusion}
In this paper, we presented \framework, a framework that supports fast
and space-efficient parallel HAC algorithms based on the
nearest-neighbor chain method. We introduced two key optimizations for
efficiency, a range query optimization and a caching optimization.
Using \framework, we designed new parallel HAC algorithms for
complete, average, and Ward linkage that outperform existing parallel
implementations by 5.8--110.1x, while using up to 237.3x less space.
It would be interesting
future work to study how to improve the efficiency of \framework by
allowing approximation.